\pdfoutput=1
\documentclass[nofootinbib, twocolumn, aps, prd]{revtex4-1}

\usepackage{graphicx}
\usepackage{natbib}
\usepackage{color}

\newcommand{\beq}{\begin{equation}}
\newcommand{\beqa}{\begin{eqnarray}}
\newcommand{\eeq}{\end{equation}}
\newcommand{\eeqa}{\end{eqnarray}}

\newcommand{\simgt}{\lower.5ex\hbox{$\; \buildrel > \over \sim \;$}}
\newcommand{\simlt}{\lower.5ex\hbox{$\; \buildrel < \over \sim \;$}}

\newcommand{\bd}[1]{\mbox{\boldmath $#1$}}

\usepackage[usenames,dvipsnames]{xcolor}

\begin{document}
\title{
Large-scale clustering as a probe of the origin
and the host environment \\of fast radio bursts
}

\author{Masato Shirasaki}
\affiliation{
National Astronomical Observatory of Japan, 
Mitaka, Tokyo 181-8588, Japan}
\email{masato.shirasaki@nao.ac.jp}

\author{Kazumi Kashiyama}
\affiliation{
Department of Physics, University of Tokyo, Tokyo 113-0033, Japan
}
\email{kashiyama@phys.s.u-tokyo.ac.jp}

\author{Naoki Yoshida}
\affiliation{
Department of Physics, University of Tokyo, Tokyo 113-0033, Japan\\
Kavli Institute for the Physics and Mathematics of the Universe (WPI),
University of Tokyo, Kashiwa, Chiba 277-8583, Japan
}
\affiliation{
CREST, Japan Science and Technology Agency, 4-1-8 Honcho, Kawaguchi, Saitama, 332-0012, Japan
}
\email{naoki.yoshida@ipmu.jp}

\begin{abstract}
We propose to use degree-scale angular clustering 
of fast radio bursts (FRBs) to identify their origin 
and the host galaxy population.
We study the information content 
in autocorrelation of the angular positions and dispersion measures (DM)
and in cross-correlation with galaxies. 
We show that the cross-correlation with Sloan Digital Sky Survey (SDSS) galaxies will place stringent constraints on 
the mean physical quantities associated with FRBs.
If $\sim$10,000 FRBs are detected 
with $\lesssim \rm deg$ resolution in the SDSS field, 
the clustering analysis 
with the intrinsic DM scatter of $100\, {\rm pc}/{\rm cm}^3$
can constrain 
the global abundance of free electrons at $z\simlt1$
and 
the large-scale bias of FRB host galaxies
(the statistical relation between the distribution of host galaxies 
and cosmic matter density field)
with fractional errors (with a $68\%$ confidence level) of 
$\sim10\%$ and $\sim20 \%$, respectively. 
The mean near-source dispersion measure and
the delay time distribution of FRB rates relative to the global
star forming rate can be also determined 
by combining the clustering and the probability distribution function of DM.
Our approach will be complementary to high-resolution ($\ll {\rm deg}$) event localization using e.g., VLA and VLBI for identifying the origin of FRBs and the source environment.
We strongly encourage future observational programs such as CHIME, UTMOST, and
HIRAX to survey FRBs in the SDSS field.
\end{abstract}

\maketitle

\section{\label{sec:intro}INTRODUCTION}

Fast radio bursts (FRBs) are millisecond transients at $\sim$ GHz frequencies characterized by their large dispersion measure (DM) of 
an order of $1000 \ {\rm pc} \, {\rm cm}^{-3}$
\cite{Lorimer+07,Keane+12,Thornton+13,Burke-Spolaor&Bannister14,Spitler+14,Petroff+15,Ravi+15,Masui+15,Champion+15}.
If the DMs are mainly due to intergalactic propagation~\cite{Ioka:2003fr, Inoue:2003ga}, FRBs are cosmological events at redshifts of $0.3-1.3$. 
Although various models have been proposed, e.g., \cite{Popov&Postnov10,Keane+12,Connor+16,Lyuvarsky14,Kashiyama+13,Totani13}, 
the origin is still uncertain.    

Recently, Refs~\cite{Chatterjee:2017fk,Marcote:2017vn} succeeded in localizing 
a repeating FRB 121102 with a submilliarcsecond resolution 
using the Karl G. Jansky Very Large Array (VLA) and the European Very Long Baseline Array Interferometry (VLBI).
The host galaxy was identified as a dwarf star-forming galaxy at $z = 0.19$~\cite{Tendulkar_et_al_17}, confirming that the FRB source is at a cosmological distance. 
Furthermore, a possible persistent radio counterpart was identified for FRB 121102~\citep{Chatterjee:2017fk,Marcote:2017vn}. 
Such a precise localization is a direct way to probe the physical properties of FRB sources and their environment and will be effective especially for repeating bursts. 
As for nonrepeating FRB, a blind survey using VLA 
will be time consuming to localize one event (see Ref~\cite{Law+15}).

Upcoming FRB surveys with e.g., CHIME, UTMOST \cite{Caleb:2016xlz}, and HIRAX \cite{Newburgh:2016mwi} will be able to detect $\sim 10,000$ FRBs per decade. 
Although the host galaxies cannot be directly identified
with the $\sim$ arcmin angular resolution, 
such a large number of FRBs can be used to probe the global abundance and spatial distribution of missing baryons \cite{Ioka:2003fr, Inoue:2003ga, McQuinn:2013tmc, Fujita:2016yve}, 
physical properties of intergalactic medium (IGM) \cite{Zheng:2014rpa},
and three-dimensional clustering of large-scale structure~\cite{Masui:2015ola}.  
It is important to develop frameworks for statistical analyses. 

In this paper, we propose to use large-scale ($\sim \rm deg$) clustering of FRBs
to study the statistical information of the host environment. 
In addition to autocorrelation analysis of FRB observables such as
sky locations and DMs~\cite{Masui:2015ola},
we consider cross-correlation analysis with Sloan Digital Sky Survey (SDSS) galaxies. 
By doing this, properties of FRB host galaxies, e.g., redshift distribution and clustering bias, can be statistically determined. 
Furthermore, the cross-correlation can be used to infer the mean value and scatter
of the DM contribution from FRB host galaxies, 
which can then be used to distinguish different models for FRBs. 

The rest of the paper is organized as follows.
In Sec.~\ref{sec:FRB}, 
we summarize FRB observables and 
their possible clustering properties.
We present a theoretical model of 
the FRB autocorrelation in Sec.~\ref{subsec:FRB-DM},
and
the cross-correlation with galaxies in Sec.~\ref{subsec:FRB-galaxy}. 
The expected signal-to-noise ratio (S/N) of the correlations are derived in Sec.~\ref{subsec:SNR}.
In Sec.~\ref{sec:Fisher}, 
we perform a Fisher analysis to study possible constraints obtained from the clustering analyses.
We also study how the constraints can be improved by combining another statistic of FRBs, i.e., probability distribution function of DM, in Sec.~\ref{subsec:DM_dist}. 
Concluding remarks and discussions are given in Sec.~\ref{sec:con}. 
Throughout the paper, we adopt the standard $\Lambda$CDM model
with the following parameters;
$\Omega_{\rm m0}=0.315$, 
$\Omega_{\Lambda}=0.685$, 
$\sigma_{8}=0.831$,
 $w_{0} = -1$,
$h=0.672$ and 
$n_s=0.964$, 
which are consistent with the PLANCK 2015 results \cite{Ade:2015xua}.

\section{\label{sec:FRB}LARGE-SCALE CLUSTERING}
\subsection{\label{subsec:obs}FRB Observables}

In this paper, we consider the dispersion measure DM$_{\rm obs}$ and angular position $\bd \theta$ as observables of FRB\footnote{
We note that the polarization and pulse profile are also important observables, from which the magnetic field and turbulent motion of gas in the line-of-sight can be inferred, respectively~\cite{Masui+15}. 
To this end, however, the signal-to-noise ratio (S/N) of the FRB should be high, and detection rate of such events will be limited.}.

\subsubsection*{Angular number density of sources}
For a given three-dimensional source distribution $n_s(\theta, z)$, 
the angular number density $n_{s, \rm 2D}(\theta)$ can be computed as
\beqa
n_{s, {\rm 2D}}({\bd \theta}) 
&=& \int_{0}^{\infty} {\rm d}z\, 
\frac{\chi^2(z)}{H(z)(1+z)} \, n_{s}({\bd \theta}, z),
\label{eq:ns2D}
\eeqa
where $z$ is the redshift, $\chi(z)$ is the comoving distance,
and $H(z)$ is the Hubble parameter. The factor of $1+z$ 
in Eq.~(\ref{eq:ns2D}) accounts 
for the effect of cosmological time dilation.
The average projected number density is then given by
\beqa
{\bar n}_{s, {\rm 2D}}
&=& \int_{0}^{\infty} {\rm d}z\, 
\frac{\chi^2(z)}{H(z)(1+z)} \, {\bar n}_{s}(z),
\label{eq:bar_ns2D}
\eeqa
where ${\bar n}_{s}(z)$ is the average comovimg number density of sources at the redshift of $z$.
From Eqs.~(\ref{eq:ns2D}) and (\ref{eq:bar_ns2D}),
one can define the angular over-density field as
\beqa
\delta_{s, {\rm 2D}}({\bd \theta}) 
&\equiv& \frac{n_{s, {\rm 2D}}({\bd \theta}) }{{\bar n}_{s, {\rm 2D}}}-1, \nonumber \\
&=& \int_{0}^{\infty} {\rm d}z\, W_{s}(z)\, \delta_{s}({\bd \theta}, z),
\label{eq:deltas_2D}
\\
W_{s}(z) 
&=& \frac{1}{{\bar n}_{s, {\rm 2D}}}
\frac{\chi^2(z)}{H(z)(1+z)}\, {\bar n}_{s}(z). \label{eq:window_s}
\eeqa
As a fiducial model, we assume that ${\bar n}_{s}(z)$ follows the star-formation history $\dot{\rho}_{*}(z)$ as
\beqa
{\bar n}_{s}(z) = {\cal A}\dot{\rho}_{*}(z)\,
\exp\left[-\frac{d_{L}^{2}(z)}{2d_{L}^{2}(z_{\rm cut})}\right],
\label{eq:ps_z}
\eeqa
where
$d_{L}(z)$ is the luminosity distance, 
the exponential form represents
an instrumental S/N threshold, and 
$\cal A$ is determined by the normalization of Eq.~(\ref{eq:bar_ns2D}). 
The star-formation history can be parametrized as 
\cite{Cole:2000ea, Hopkins:2006bw}
\beqa
\dot{\rho}_{*}(z) \propto \frac{{\alpha}_{0}+{\alpha}_{1}z}{1+(z/{\alpha}_{2})^{{\alpha}_3}},
\label{eq:SFH}
\eeqa
with ${\alpha}_{0}=0.0170$, ${\alpha}_1 = 0.13$, ${\alpha}_2=3.3$, and ${\alpha}_3=5.3$. 
Our fiducial model [Eq.~(\ref{eq:ps_z})] is consistent with
an estimated redshift distribution of the observed FRBs \cite{Petroff:2016tcr}
if the redshift cutoff is set to be $z_{\rm cut} = 0.5$ \cite{Munoz:2016tmg}.
Note that our results are less sensitive to 
${\alpha}_{2}$ and ${\alpha}_{3}$ 
since these parameters determine the redshift distribution at $z\simgt2$.
The dependence of $z_{\rm cut}$ on the clustering analysis is summarized in Sec.~\ref{subsec:SNR}. There, we found that the choice of $z_{\rm cut}$ have a small impact on the signal-to-noise in autocorrelation of DM and the cross correlation of DM and galaxies.
In Sec.~\ref{sec:Fisher}, we examine another model of ${\bar n}_{s}$ taking into account a time delay of FRB rates relative to the global star-forming rate [see Eq.~(\ref{eq:delay_model})].

\subsubsection*{Two-dimensional field of dispersion measures}
${\rm DM}_{\rm obs}({\bd \theta})$ is defined as the integral of number density of free electrons along a line of sight,
which can be decomposed as
\beqa
{\rm DM}_{\rm obs}
= {\rm DM}_{\rm IGM}+{\rm DM}_{\rm host}
+{\rm DM}_{\rm MW}, 
\eeqa
where ${\rm DM}_{\rm IGM}$, ${\rm DM}_{\rm host}$
and ${\rm DM}_{\rm MW}$ represent the contributions 
from the IGM, FRB host galaxies, and the Milky way, respectively.
${\rm DM}_{\rm host}$ includes the interstellar medium of the host and near-source plasma.
We assume that ${\rm DM}_{\rm MW} ({\bd \theta})$ for each direction is already determined from Galactic pulsar observations~\cite{Taylor:1993my} and can be subtracted from ${\rm DM}_{\rm obs} ({\bd \theta})$\footnote{
In real, the subtraction of ${\rm DM}_{\rm MW}$ is 
still uncertain and the imperfect subtraction can affect the measurement of 
the autocorrelation of DM. 
On the other hand, the cross-correlation analysis of DM with extragalactic objects should be insensitive to the subtraction of galactic DM,
since the galactic DM does not correlate with the spatial distributions of extragalactic objects.
As we show in the following sections, the cross correlation of DM and 
galaxies can play a central role to constrain the parameters of FRB sources,
indicating that the imperfect subtraction will not affect our results significantly.}. 
In the following, we focus on the extragalactic DM field expressed as ${\rm DM}_{\rm ext} = {\rm DM}_{\rm IGM} +  {\rm DM}_{\rm host}$.

For a fixed source redshift $z_{s}$, ${\rm DM}_{\rm IGM}$ is given by
\beqa
{\rm DM}_{\rm IGM}({\bd \theta},z_{s})
= \int_{0}^{z_{s}}\frac{{\rm d}z}{H(z)}
\frac{n_{e}({\bd \theta}, z)}{(1+z)^2},
\label{eq:def_DM_IGM}
\eeqa
where $n_{e}({\bd \theta}, z)$ represents the three-dimensional 
number density of free electrons at redshift $z$.
The average density of IGM elections can be expressed as
\cite{Deng:2013aga, Zheng:2014rpa}
\beqa
{\bar n}_{e}(z) &=&n_{0}(1+z)^{3}f_{e}(z)
, \label{eq:bar_ne}
\eeqa
where 
\beqa
n_{0} = \frac{\Omega_{b}\rho_{\rm crit,0}}{m_{H}} 
=2.475\times10^{-7}\left(\frac{\Omega_{b}h^2}{0.022}\right)\, 
{\rm cm}^{-3},
\eeqa
with $\Omega_{b}$ being the baryon density normalized by the critical density $\rho_{\rm crit,0}$ at $z=0$
and $h=H(z=0)/(100\, {\rm km}/{\rm s}/{\rm Mpc})$.
Also,
\beqa
f_{e} = \left[ (1-Y)f_{\rm HII}+\frac{1}{4}Y(f_{\rm HeII}+2f_{\rm HeIII})\right], 
\eeqa
where $Y\simeq0.25$ is the mass fraction of helium,
$f_{\rm HII}$ is the ionization fraction of hydrogen, 
$f_{\rm HeII}$ and $f_{\rm HeIII}$ represent the ionization fractions of singly and doubly ionized helium, respectively.
After helium reionization (occurred at $z\sim$2-3), 
we can approximate as $f_{\rm HII}=1$, $f_{\rm HeII}=0$, and $f_{\rm HeIII}=1$.
In this case, $f_{e} = 0.88$ and 
\beqa
{\rm DM}_{\rm IGM}({\bd \theta},z_{s})
&=& 1060\,{\rm pc} \ {\rm cm}^{-3}\, 
\left(\frac{f_{e}}{0.88}\right)
\left(\frac{\Omega_{b}h^2}{0.022}\right)
\left(\frac{h}{0.7}\right)^{-1} \nonumber \\
&& \times
\Bigl\{
\int_{0}^{z_{s}}{\rm d}z\, \frac{(1+z)}{E(z)} 
\left[1+\delta_{e}({\bd \theta},z_{s})\right]
\Bigr\}, \label{eq:DM_IGM_singlez}
\eeqa
where $E(z)=H(z)/H(z=0)$ and $\delta_{e}$ is the over-density field 
of free electrons.
Note that Eq~(\ref{eq:DM_IGM_singlez}) with $\delta_e=0$
correspond to Eq~(2) in Ref~\cite{Inoue:2003ga}, which is commonly used.
The average ${\rm DM}_{\rm IGM}$ for an angular position ${\bd \theta}$
can be described as 
\beqa
{\rm DM}_{\rm IGM}({\bd \theta}) 
= \int_{0}^{\infty}{\rm d}z\, W_{\rm DM, IGM}(z)
\left[1+\delta_{e}({\bd \theta},z)\right],
\label{eq:DM_IGM_variousz}
\eeqa
where 
\beqa
W_{\rm DM, IGM}(z) 
&=& 1060\,{\rm pc} \ {\rm cm}^{-3}\, 
\left(\frac{f_{e}}{0.88}\right)
\left(\frac{\Omega_{b}h^2}{0.022}\right)
\left(\frac{h}{0.7}\right)^{-1} \nonumber \\
&& \times 
\frac{(1+z)}{E(z)}
\int_{z}^{\infty}W_{s}(z) dz.
\eeqa

We next consider the contribution from host galaxies.
For galaxies at redshift of $z_s$, 
${\rm DM}_{\rm host}$ is expressed as
\beqa
{\rm DM}_{\rm host}({\bd \theta}, z_{s})
&=& \int {\rm d}^2\theta_{s}\, \tau_{e}(\chi_s\left[{\bd \theta}-{\bd \theta}_s\right]|z_{s}) \nonumber \\
&&
\times{\bar n}_{s}(z_s) \left[ 1+\delta_{s}({\bd \theta}_{s}, z_{s})\right]\label{eq:def_DM_host_singlez}
\eeqa
where $\chi_s=\chi(z_s)$ and
$\tau_{e}({\bd x}_{\perp}| z_{s})$ represent 
the projected number density of free electrons around the host galaxy
and the apparent angular size, respectively. 
In this paper,  the apparent angular size of $\tau_{e}$ is assumed to be small enough, 
i.e. $\tau_{e}({\bd x}, z) \propto \delta^{(2)}({\bd x})$,
where $\delta^{(2)}({\bd x})$ is the two-dimensional delta function.
This approximation should be reasonable when one considers the large-scale
clustering of DM with angular separation of $\simgt1$ deg.
Taking into account the source distribution, the average ${\rm DM}_{\rm host}$ for an angular position ${\bd \theta}$ is given by
\beqa
{\rm DM}_{\rm host}(\bd \theta)
= \int_{0}^{\infty}{\rm d}z\, W_{\rm DM, host}(z)
\left[ 1+\delta_{s}({\bd \theta}, z) \right],
\label{eq:DM_host_variousz}
\eeqa
where 
\beqa
W_{\rm DM, host}(z) = {\bar \tau}_{e}(z) W_{s}(z).
\eeqa
In the above equation, ${\bar \tau}_{e}(z)$
represents the mean DM from galaxies at redshift of $z$.
Note that ${\bar \tau}_e$ is considered to be averaged over the orientation
and population of the host galaxies.
The redshift dependence of ${\bar \tau}_e$ should contain the information of the environment of FRB sources, which is poorly known.
In this paper, we assume ${\bar \tau}_{e}(z)$ to be constant, for simplicity. 
We take ${\bar \tau}_{e} = 100 \ \rm pc \ cm^{-3}$ as a face value, 
that is consistent with the observational constraint on the host galaxy of
FRB 121102~\cite{Tendulkar_et_al_17}. 

According to Eqs~(\ref{eq:DM_IGM_variousz}) and (\ref{eq:DM_host_variousz}), 
${\rm DM}_{\rm IGM}$ and 
${\rm DM}_{\rm host}$
can be expressed as the integral of 
over-density field of electron number density $\delta_{e}$
and source number density $\delta_{s}$ along a line of sight, respectively.
In order to compute a possible clustering signal of these DMs,
we adopt the linear bias model.
In the linear bias model, a given over-density field $\delta_{\alpha}$ is
expressed as
\beqa
\delta_{\alpha}(\bd{x}) &=& b_{\alpha} \delta_{m} (\bd{x}),
\eeqa
where $\delta_{m}$ represents the overndensity of cosmic 
matter density.
In this model, the distribution of $\delta_{\alpha}$ 
can be determined by $\delta_{m}$, but the amplitude of their fluctuation
is {\it biased} by a factor of $b_{\alpha}$.
The proportional factor $b_{\alpha}$ is referred to as 
the bias factor throughout this paper.
The linear bias model is thought to be valid for the clustering analysis on large scales greater than $\sim10\, {\rm Mpc}$ \cite{1999MNRAS.308..119S}.

\subsection{\label{subsec:FRB-DM}FRB autocorrelation}

We then consider the large-scale clustering of FRBs.
In general, clustering information of a two-dimensional field $f({\bd \theta})$ is encompassed in the two-point correlation function; 
\beqa
\xi_{ff}(\theta) = 
\langle f({\bd \phi})f({\bd \phi}+{\bd \theta})\rangle
-\langle f({\bd \phi})\rangle
\langle f({\bd \phi}+{\bd \theta})\rangle.
\eeqa
The power spectrum $C_{ff}(\ell)$ defined as  
\beqa
C_{ff}(\ell)=\int {\rm d}^2\theta\, \xi_{ff}(\theta)\exp
\left(-i{\bd \ell}\cdot{\bd \theta}\right),
\eeqa
is commonly used in clustering analyses.
Here $\ell=2\pi/\theta$ is the multipole. 
In this paper, we adapt the flat-sky approximation. 

Using Eqs.~(\ref{eq:deltas_2D}) and (\ref{eq:window_s}) with the Limber approximation \cite{Limber:1954zz},
we can compute the angular power spectrum of the over-density field of FRB sources as
\beqa
C_{ss}(\ell) 
&=& 
\int {\rm d}z\, W_{s}^2(z) \frac{H(z)}{\chi^2(z)} \nonumber \\
&&\, \times \,
b_{\rm FRB}^2 P_{m}\left(\frac{\ell+1/2}{\chi}, z\right),
\eeqa
where $b_{\rm FRB}$ is the bias factor of $\delta_s$ 
relative to the underlying matter over-density field $\delta_m$
and $P_m(k,z)$ represents the three-dimensional power spectrum of 
$\delta_m$ at redshift $z$. 
We assume that $P_m(k,z)$ is the linear matter power spectrum. 
The approximation of using linear matter power spectrum
is valid at sufficiently large scales of $k\simlt0.1 h/{\rm Mpc}$.
The linear matter power spectrum is computed with {\tt CAMB} \cite{Lewis:1999bs}.
We also assume linear bias of $\delta_{s}({\bd \theta}, z)=b_{\rm FRB} \delta_{m}({\bd \theta}, z)$ 
and compute the angular power spectrum of the extragalactic DM field ${\rm DM}_{\rm ext}$ as
\beqa
\label{eq:C_DMDM}
C_{\rm DM-DM}(\ell)
&=&C_{\rm IGM-IGM}(\ell)+C_{\rm IGM-host}(\ell)\nonumber \\
&&
\,\,\,\,\,\,\,\,\,\
\,\,\,\,\,\,\,\,\,\
\,\,\,\,\,\,\,\,\,\
\,\,\,\,
+C_{\rm host-host}(\ell), \\
C_{\rm IGM-IGM}(\ell)
&=&
\int {\rm d}z\, W_{\rm DM, IGM}^2(z) \frac{H(z)}{\chi^2(z)} \nonumber \\
&&\, \times \,
b_{e}^2 P_{m}\left(\frac{\ell+1/2}{\chi}, z\right), \\
C_{\rm IGM-host}(\ell)
&=&
\int {\rm d}z\, 2W_{\rm DM, IGM}(z) W_{\rm DM, host}(z)
\frac{H(z)}{\chi^2(z)} \nonumber \\
&&\, \times \,
b_{e} b_{\rm FRB} P_{m}\left(\frac{\ell+1/2}{\chi}, z\right), \\
C_{\rm host-host}(\ell)
&=&
\int {\rm d}z\, W_{\rm DM, host}^2(z) \frac{H(z)}{\chi^2(z)} \nonumber \\
&&\, \times \,
b_{\rm FRB}^2 P_{m}\left(\frac{\ell+1/2}{\chi}, z\right),
\eeqa
where $b_{e} = \delta_{e}/\delta_m$ is the bias factor of electron density field. 

\begin{figure}[!t]
\begin{center}
       \includegraphics[clip, width=0.9\columnwidth, bb=0 0 438 430]
       {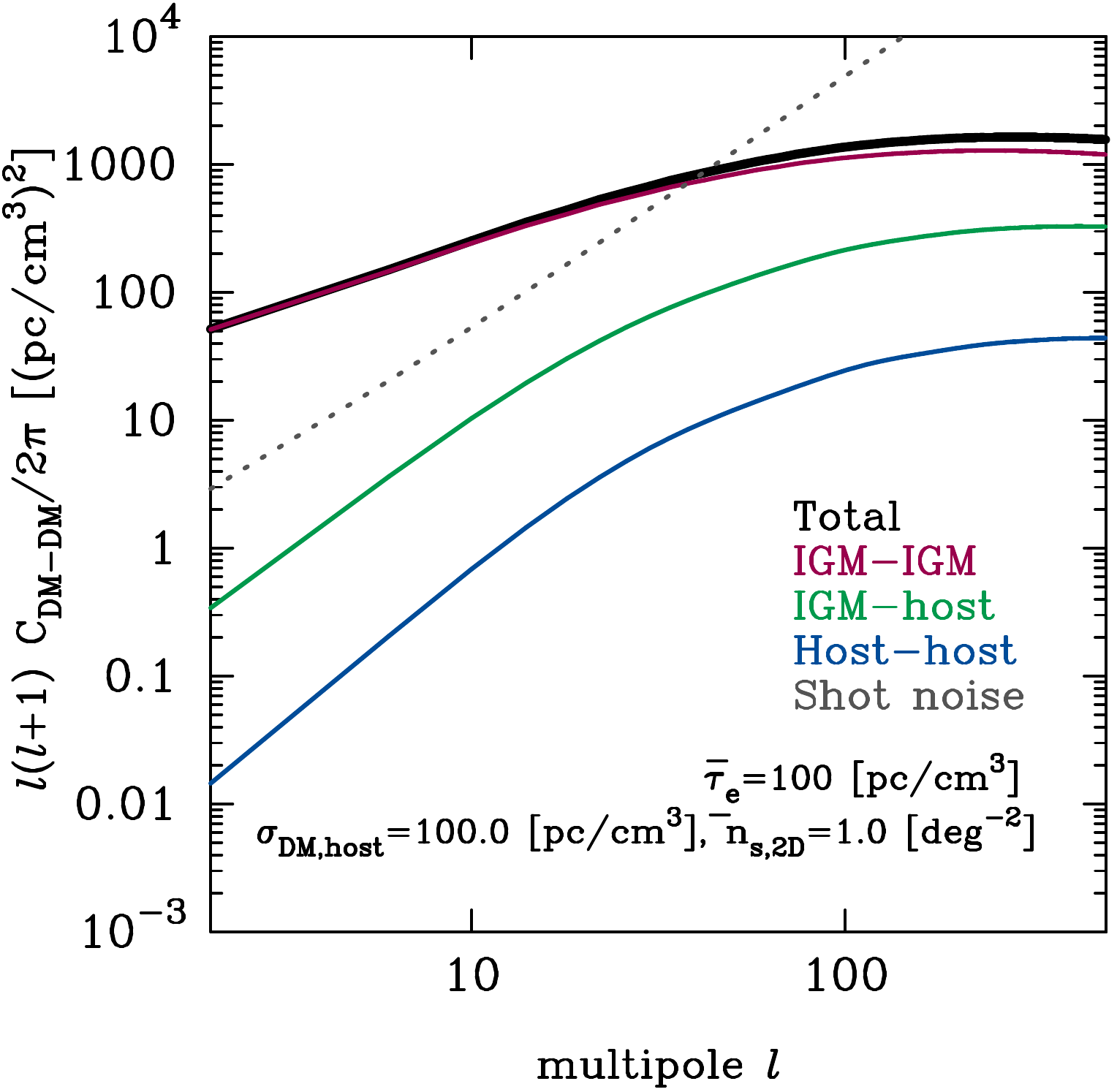}
     \caption{
     \label{fig:DM_llcl}
     Auto power spectrum of dispersion measure (DM).
     Colored lines represent contributions from 
     the autocorrelation of the IGM component (red), 
     the cross-correlation of the IGM and host-galaxy components (green), 
     and autocorrelation of the host-galaxy component (blue). 
     The solid black line shows the total power. 
     The mean DM around host galaxies ${\bar \tau}_{e}$ is set to 
     be $100\, {\rm pc} \ {\rm cm}^{-3}$.
     The dashed gray line indicates the shot noise induced by the intrinsic scatter
     of DM around host galaxies $\sigma_{\rm DM, host}$.
     Here we assume the average source number density of $\bar n_{\rm s, 2D} = 1 \ {\rm deg}^{-2}$ and $\sigma_{\rm DM, host}=100\, {\rm pc} \ {\rm cm}^{-3}$.
     }
    \end{center}
\end{figure}

Fig.~\ref{fig:DM_llcl} shows the auto power spectrum of DMs, $C_{\rm DM-DM}$.
Here we set $b_{\rm FRB}=1.3$, $b_{e}=1$,
and ${\bar \tau}_{e}=100\, {\rm pc} \ {\rm cm}^{-3}$.
Note that $b_{\rm FRB}=1.3$ is consistent with 
star-forming galaxies at $z \lesssim1$ \cite{Blake:2011rj}.
In this case, the clustering of IGM will dominate. 
We find that $C_{\rm IGM-host}$ becomes larger than $C_{\rm IGM-IGM}$ at $\ell = 100$ if $b_{\rm FRB}{\bar \tau}_e \simgt680 \, {\rm pc} \ {\rm cm}^{-3}$. 
The dashed line in Fig.~\ref{fig:DM_llcl} represents the shot noise
induced by the intrinsic scatter of DM around host galaxies $\sigma_{\rm DM, host}$.
The shot noise $N_{\rm DM-DM}$ is computed as
\beqa
N_{\rm DM-DM} 
&=& \frac{\sigma_{\rm DM, host}^2}{{\bar n}_{s, {\rm 2D}}} \\
&=& 1.95\,  \left({\rm pc} \ {\rm cm^{-3}}\right)^2
\left(\frac{\sigma_{\rm DM, host}}{100\, {\rm pc} \ {\rm cm^{-3}}}\right)^2 \nonumber \\
&&\times
\left(\frac{{\bar n}_{s, {\rm 2D}}}{1\, {\rm deg}^{-2}}\right)^{-1}.
\eeqa
With a FRB number density of $\bar n_{\rm 2D, s} \gtrsim 1 \ \rm deg^{-2}$, 
the signal $C_{\rm DM-DM}$ is larger than the noise $N_{\rm DM-DM}$ at $\ell \lesssim 100$ if $\sigma_{\rm DM, host} \lesssim 100 \ \rm pc \ cm^{-3}$. 
We study the information content of $C_{\rm DM-DM}$ in more detail
in Sec.~\ref{subsec:SNR}.
 
\subsection{\label{subsec:FRB-galaxy}Cross-correlation with galaxies}

The clustering analysis of the FRB autocorrelation can give some constraints on the model parameters of FRB sources and IGM,
but they will be degenerate. 
We here consider cross-correlation analysis with galaxies in order to put additional constraints. 

In general, galaxies trace the large-scale structure in a biased manner.  
The bias factor depends on the type of galaxies.  
Thus, the host galaxies of FRBs and their redshift evolution can be statistically inferred from spatial cross-correlation between FRBs and galaxies. 
A similar idea has been proposed in Ref~\cite{Oguri:2016dgk} 
to constrain the redshift-distance relation of gravitational-wave sources.
In principle, the three-dimensional information 
can be extracted from observables
of FRBs alone; 
Ref~\cite{Masui:2015ola} proposes 
that DMs can be used as the distance indicator as similar to redshift.

\begin{table}
\begin{center}
\begin{tabular}{|c|c|c|c|}
\tableline
Sample & Redshift range & ${\bar n}_{g}\, (h/{\rm Mpc})^3$ 
& Galaxy bias $b_g$ \\ \tableline
LOW-Z & $0.15<z<0.43$ & $3\times10^{-4}$  & 1.7 \\
CMASS & $0.43<z<0.70$ & $3\times10^{-4}$  & 1.9 \\ 
eBOSS & $0.70<z<1.60$ & $3\times10^{-4}$   & 1.3 \\
\tableline
\end{tabular} 
\caption{
\label{tab:gal}
Summary of galaxy samples assumed in this paper.
The galaxy bias of LOW-Z and CMASS are found 
to be consistent with the previous works in 
Refs~\cite{Manera:2012sc, Manera:2014cpa}.
}
\end{center}
\end{table}

\begin{figure*}
\begin{center}
       \includegraphics[clip, width=1.50\columnwidth, bb=0 20 520 256]
       {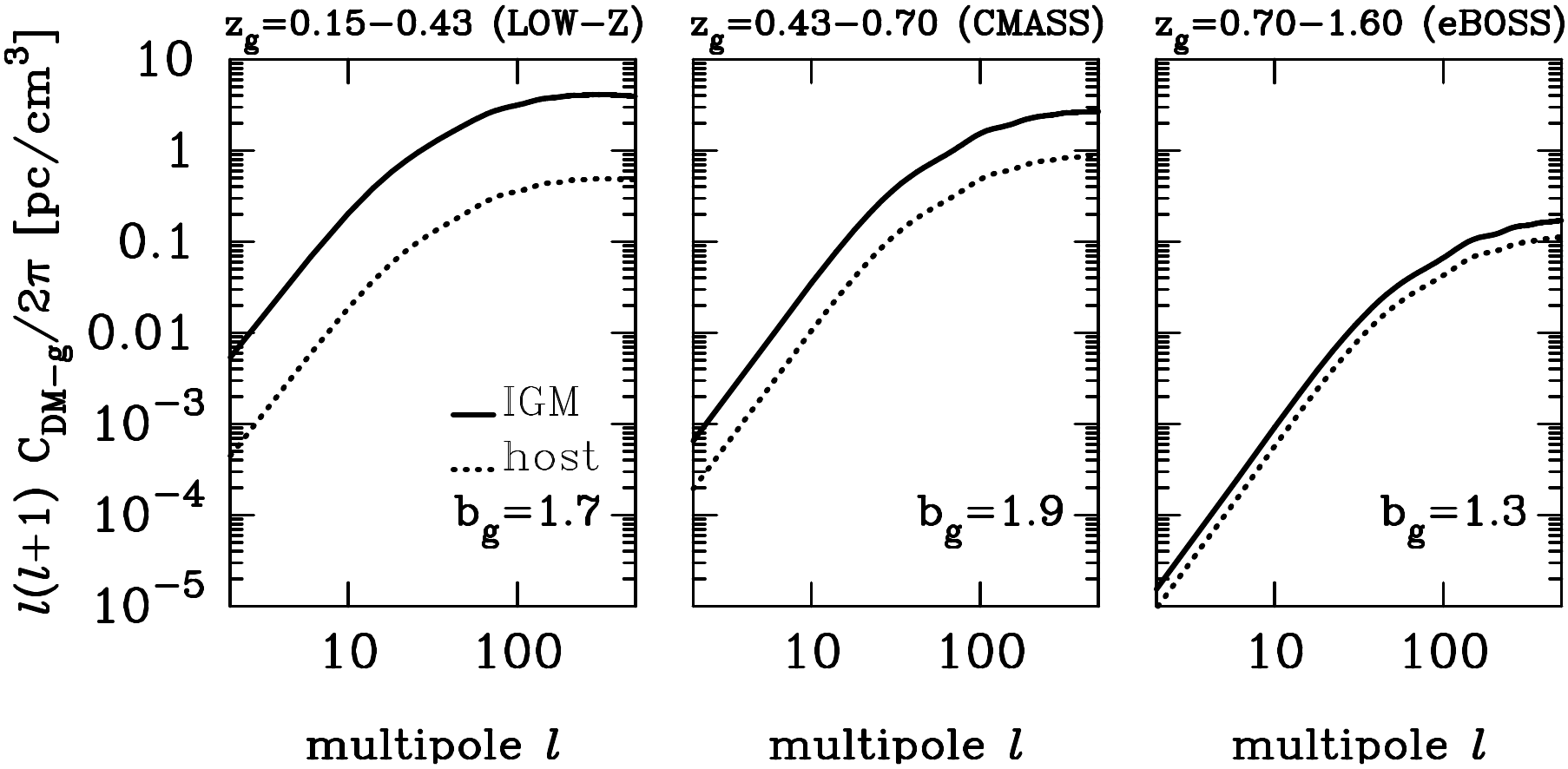}
      \includegraphics[clip, width=1.50\columnwidth, bb=0 0 520 256]
      {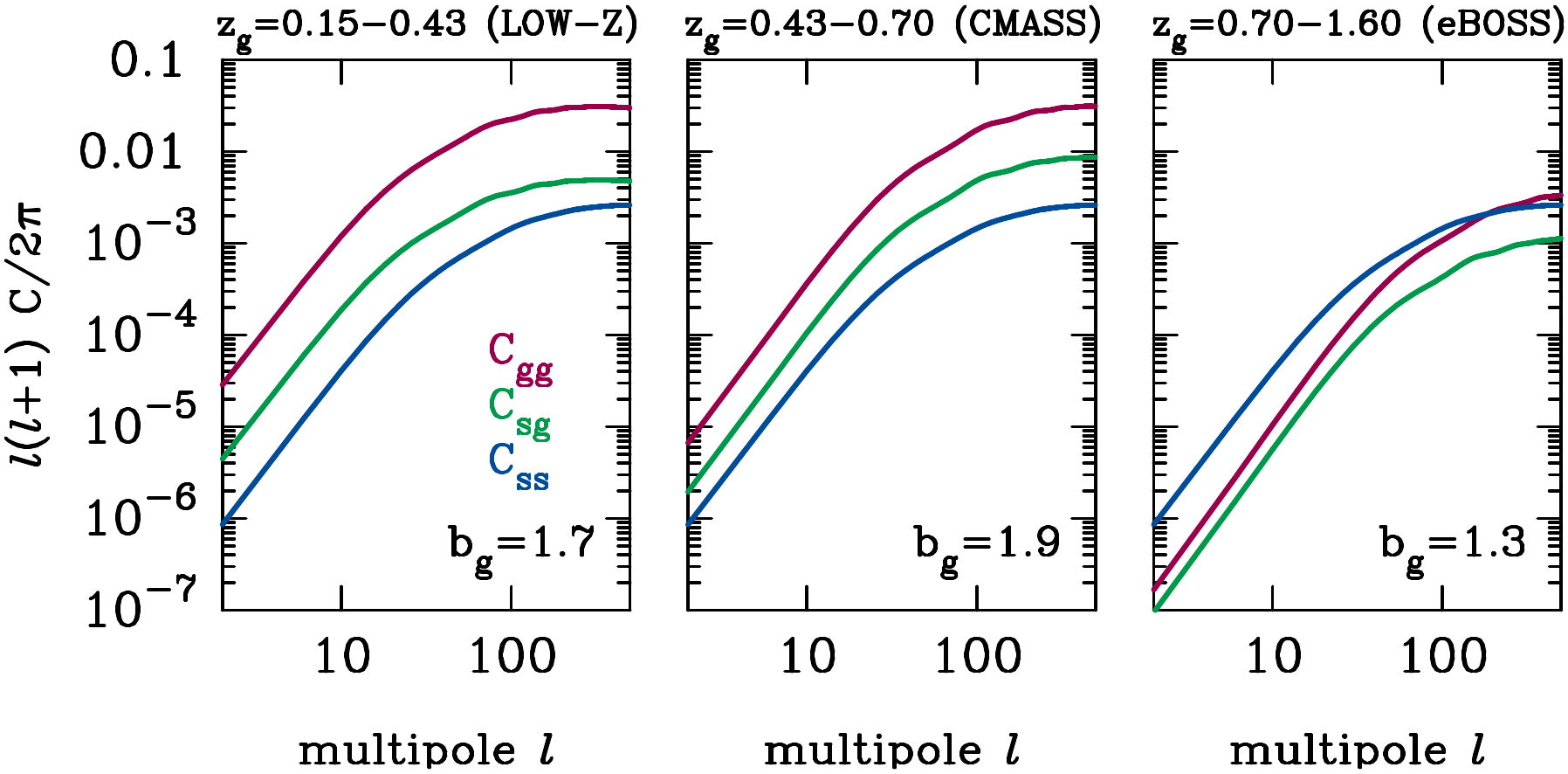}
     \caption{
     \label{fig:cross_llcl}
     Cross power spectra of FRB observables and galaxies.
     The upper panels show the cases for dispersion measures.
     The solid and dotted lines correspond to the contribution from the IGM and host galaxies, respectively.  
     The bottom panels show the cases for angular number density of FRBs (green lines). 
     We also show the auto power spectra of FRBs (blue lines) and galaxies (red lines) for comparison.  
     Note that the similar shape among cross power spectra is expected in the linear bias model. In the linear bias model, the distribution of the relevant field such as electron number density is assumed to follow the matter density field.
     }
    \end{center}
\end{figure*}

Let us consider a spectroscopic sample of galaxies with redshift ranging from $z_{\rm i, min}<z<z_{\rm i, max}$.
The over-density field of galaxies is expressed  
[in a similar way to Eq.~(\ref{eq:deltas_2D})] as
\beqa
\delta^{i}_{g, {\rm 2D}}(\bd \theta)
= \int_{0}^{\infty} {\rm d}z\, W_{g, i}(z) \delta_{g, i}({\bd \theta},z), 
\eeqa
where $\delta_{g, i}$ represents the three-dimensional over-density field of galaxies. 
The window function $W_{g,i}(z)$ is 
\beqa
W_{g,i}(z) 
&=&
\frac{1}{{\bar n}^{i}_{g, {\rm 2D}}}
\frac{\chi^2}{H(z)}{\bar n}_{g}(z) \nonumber \\
&&\times
{\cal H}(z-z_{\rm i, max})
{\cal H}(z_{\rm i, min}-z),
\eeqa
where ${\bar n}_{g}(z)$ is the average comoving number density of galaxies, 
${\cal H}(x)$ is the Heaviside function, and ${\bar n}^{i}_{g, {\rm 2D}} = \int_{z_{\rm i, min}}^{z_{\rm i, max}} {\rm d}z\, \chi^2 \bar{n}_{g}(z)/H(z)$.

Using the Limber approximation \cite{Limber:1954zz}, 
the cross power spectrum of $\delta^{i}_{{g,\rm 2D}}$ and $\delta_{s, \rm 2D}$ can be given as 
\beqa
C_{sg,i}(\ell)
&=& \int {\rm d}z\, W_{s}(z)W_{g,i}(z) \frac{H(z)}{\chi^2(z)} \nonumber \\
&&\, \times \,
b_{\rm FRB}b_{g, i} P_{m}\left(\frac{\ell+1/2}{\chi}, z\right),
\label{eq:C_sg}
\eeqa
where we assume linear bias of 
$\delta_{g,i}({\bd \theta},z) = b_{g,i}\delta_m({\bd \theta}, z)$.
For each galaxy sample (identified by the index i), 
the correlation arises from the clustering in a finite redshift range of 
$z_{\rm i, min}<z<z_{\rm i, max}$.
Therefore, $C_{sg,i}(\ell)$ contains the information of the source distribution $W_s(z)$.
We can also compute the cross power spectrum
of $\delta^{i}_{{g,\rm 2D}}$ and ${\rm DM}_{\rm ext}$ as
\beqa
C_{{\rm DM}-g,i}
&=& C_{{\rm IGM}-g, i} + C_{{\rm host}-g,i}, 
\eeqa
where
\beqa
C_{{\rm IGM}-g,i} \label{eq:C_DMg}
&=&
\int {\rm d}z\, W_{\rm DM, IGM}(z)W_{g,i}(z) \frac{H(z)}{\chi^2(z)} 
\nonumber \\
&&\, \times \,
b_{e}b_{g, i} P_{m}\left(\frac{\ell+1/2}{\chi}, z\right), \\
C_{{\rm host}-g,i}
&=&
\int {\rm d}z\, W_{\rm DM, host}(z)W_{g,i}(z) \frac{H(z)}{\chi^2(z)} 
\nonumber \\
&&\, \times \,
b_{\rm FRB}b_{g, i} P_{m}\left(\frac{\ell+1/2}{\chi}, z\right),
\eeqa
which also contains the information of source distribution $W_{s}(z)$.  
Moreover, the mean DM from host galaxies ${\bar \tau}_e$ and the linear bias of sources $b_{\rm FRB}$ 
can be inferred from these power spectra. 

In this paper, we consider three spectroscopic samples 
of galaxies from SDSS.
These include two samples from the SDSS-III Baryon Oscillation Spectroscopic Survey (BOSS),
named LOW-Z and CMASS \cite{Reid:2015gra}.
We also consider emission-line galaxies to be catalogued
by the SDSS-IV extended Baryonic Oscillation 
Spectroscopic Survey (eBOSS) \cite{Dawson:2015wdb}.
The characteristics of these samples are summarized in Table~\ref{tab:gal}.

Fig.~\ref{fig:cross_llcl} shows the expected cross power spectra.
We assume $b_{\rm FRB}=1.3$, $b_{e}=1$, and ${\bar \tau}_{e}=100\, {\rm pc} \ {\rm cm}^{-3}$ as in Figure \ref{fig:DM_llcl}.
For both $C_{sg,i}$ and $C_{{\rm DM}-g,i}$, 
the highest redshift bin has the smallest power. 
This is because we set an exponential cutoff for the FRB source distribution as $z_{\rm cut} = 0.5$ (see Eq.~\ref{eq:ps_z}).
When the mean DM from host galaxies is set to be $100\, {\rm pc} \ {\rm cm}^{-3}$, 
the contribution from IGM is dominant in the range of $0.15<z<0.70$, 
while the contribution from host galaxies can become important at $z>0.70$.
We note that the contribution from IGM is proportional to the integration term $\int_{z}^{\infty} {\rm d}z\, W_s(z)$
that is decreasing quickly for higher redshift in the presence of the exponential cutoff as in Eq.~(\ref{eq:ps_z}).

\begin{figure}
\begin{center}
       \includegraphics[clip, width=0.90\columnwidth, bb=0 0 462 439]
       {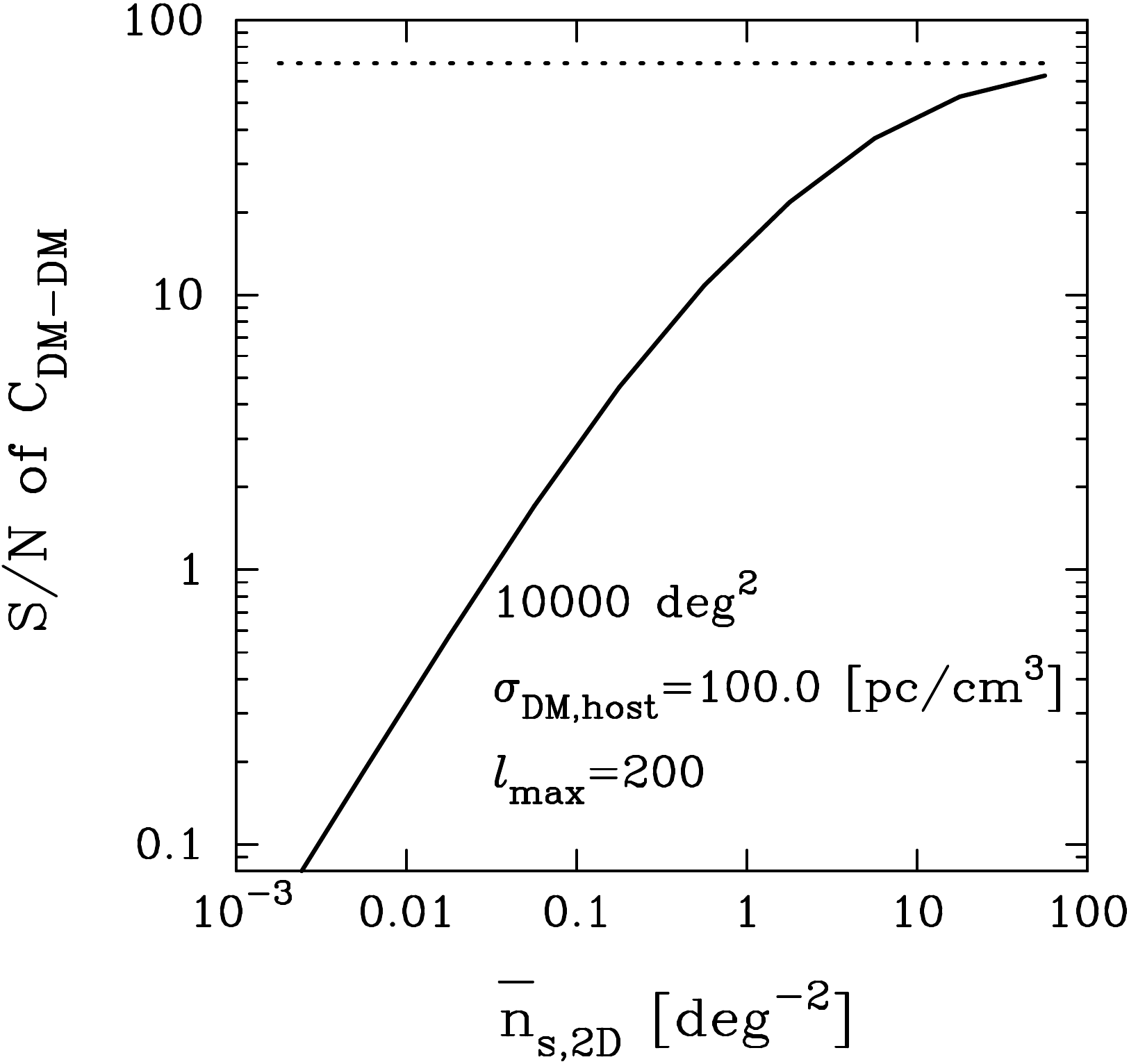}
       \includegraphics[clip, width=0.90\columnwidth, bb=0 0 452 441]
       {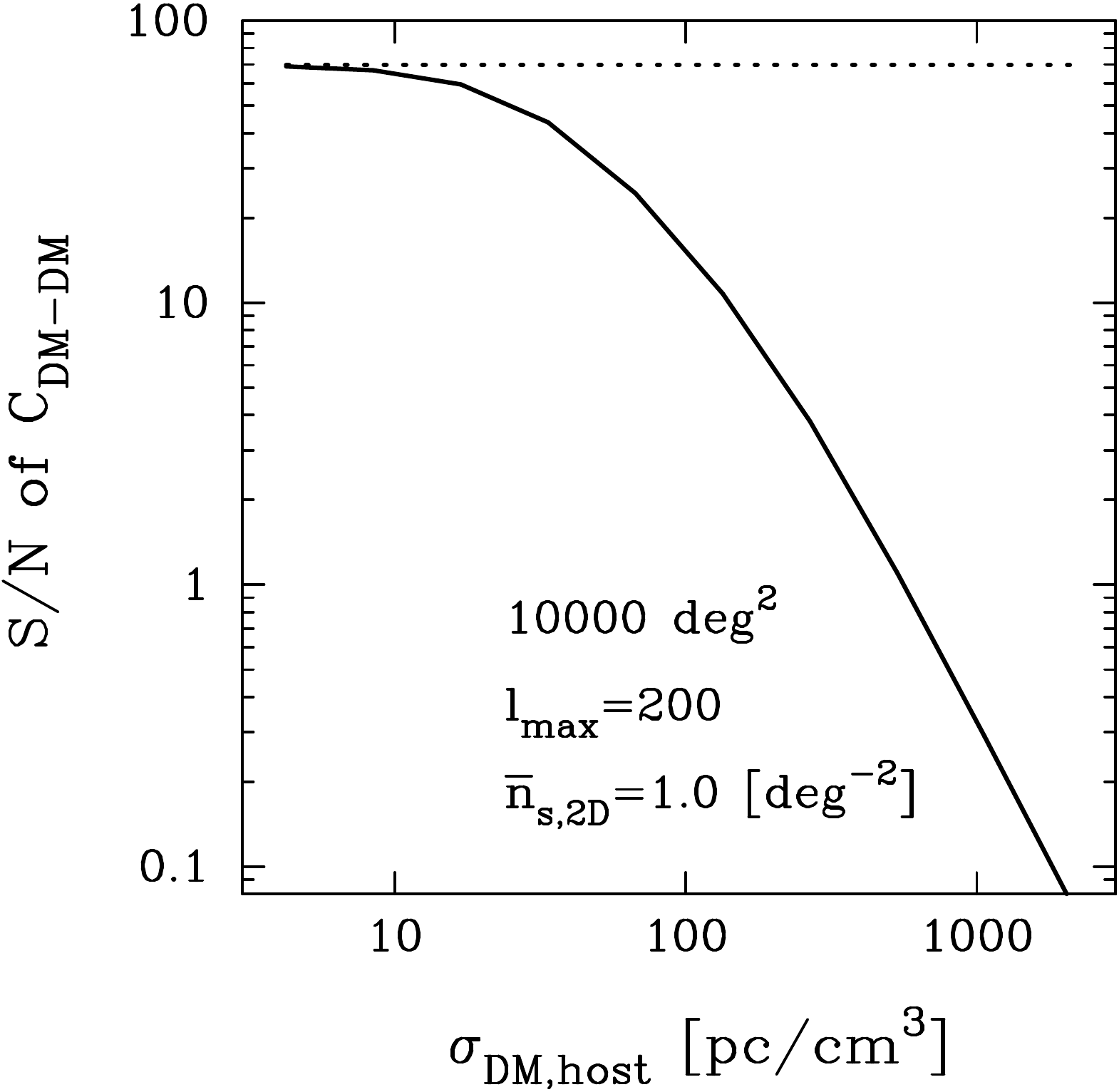}
     \caption{
     \label{fig:SN_DM_llcl}
     Signal-to-noise ratio (S/N) of the autocorrelation of DMs.
     In the top panel, we plot the S/N 
     as a function of the average source number density $\bar n_{\rm s, 2D}$
     for a fixed intrinsic scatter of DM around host galaxies, 
     $\sigma_{\rm DM, host}=100\, {\rm pc} \ {\rm cm}^{-3}$.
     In the bottom panel, we plot the S/N
     as a function of $\sigma_{\rm DM, host}$ 
     for $\bar n_{\rm s, 2D} = 1 \ {\rm deg}^{-2}$.
     In both, we set the maximum multipole to be $\ell_{\rm max}=200$
     and assume a sky coverage of 10,000 ${\rm deg}^2$.
     The dotted lines in both panels represent the case
     in the absence of the shot noise 
     (i.e., $\sigma_{\rm DM, host}=0\, {\rm pc} \ {\rm cm}^{-3}$).
     }
    \end{center}
\end{figure}

\begin{figure*}
\begin{center}
       \includegraphics[clip, width=1.50\columnwidth, bb=0 50 542 254]
       {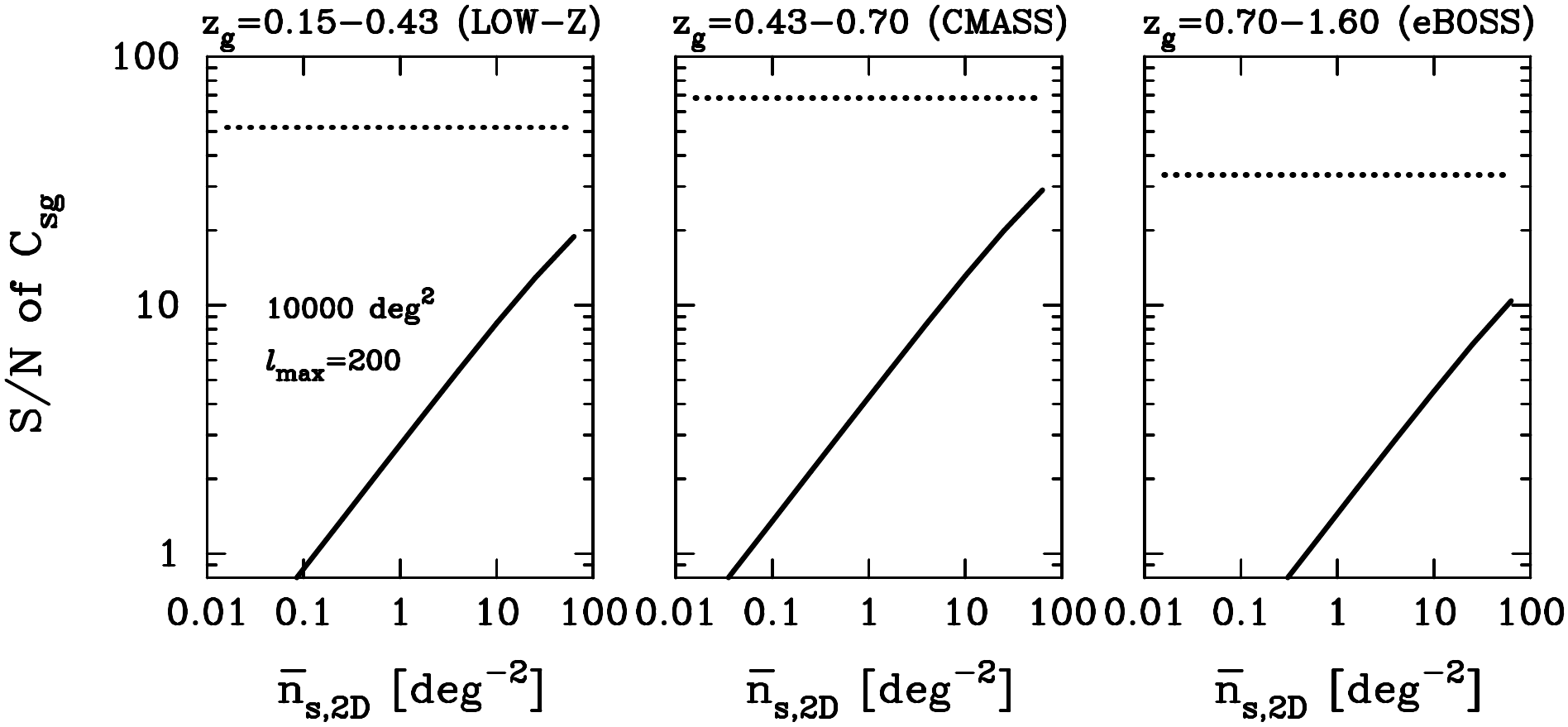}
       \includegraphics[clip, width=1.50\columnwidth, bb= 0 0 542 264]
       {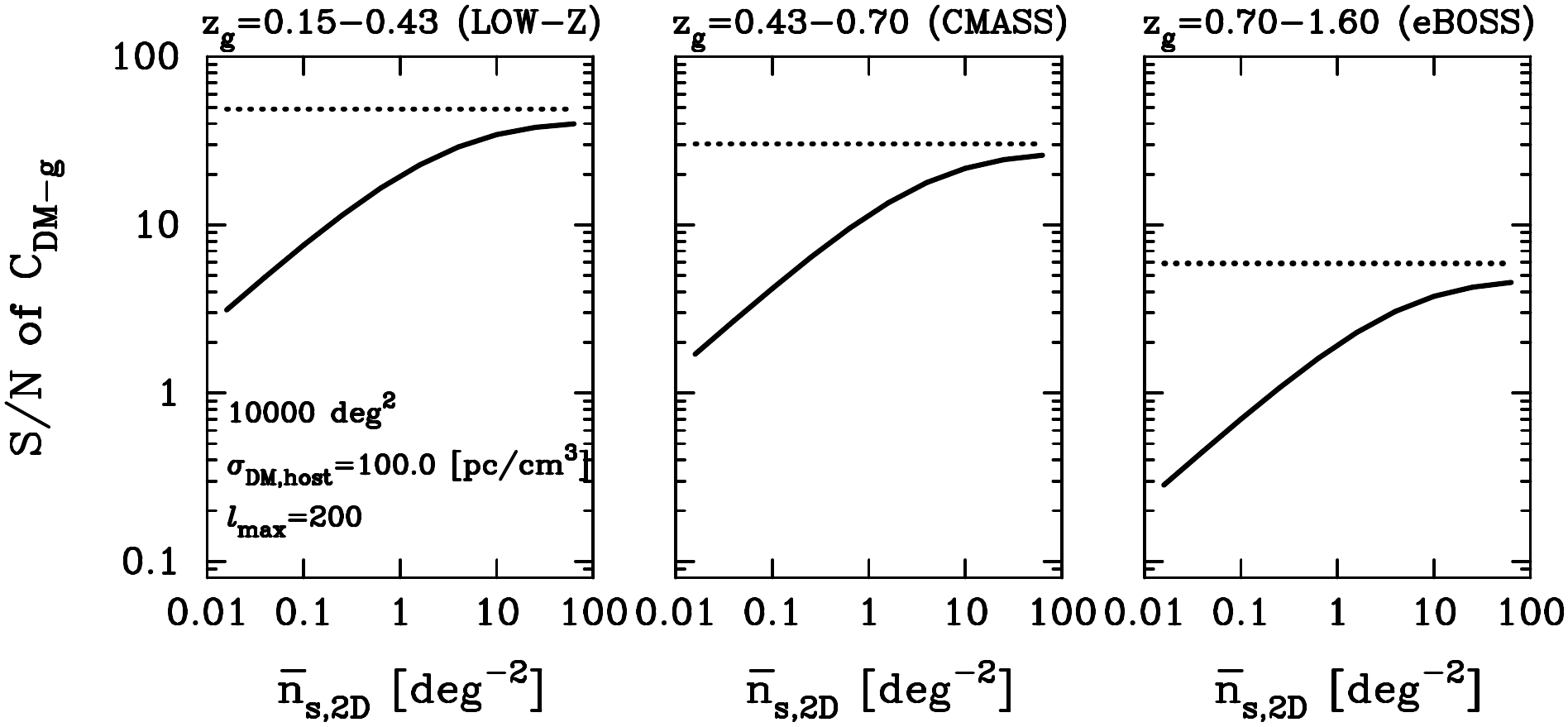}
     \caption{
     \label{fig:SN_cross_llcl}
     S/N of the cross-correlation of FRBs and three spectroscopic galaxy samples as a function of the average source number density of FRBs.
     The top and bottom panels show the S/N of $C_{sg, i}$ and $C_{{\rm DM}-g,i}$, respectively.
     In the bottom panels, we assume the intrinsic scatter of DM to be $\sigma_{\rm DM, host}= 100\,{\rm pc} \ {\rm cm}^{-3}$. 
     The dotted lines represent the case 
     with $\sigma_{\rm DM, host}=0$ and $\bar{n}_{s, \rm 2D}
     \rightarrow \infty$.
     }
    \end{center}
\end{figure*}

\subsection{\label{subsec:SNR}Signal-to-noise ratio}
The S/N of angular power spectrum essentially determines 
to what extent we can extract source information from the clustering analysis. 
For given multiple field $X$ and $Y$, the S/N
of cross power spectrum $C_{XY}(\ell)$ can be computed as
\beqa
\left[\frac{\rm S}{\rm N}\right]^2 (\ell_{\rm max})
&=& \sum_{\ell_i, \ell_j < \ell_{\rm max}}
{\rm Cov}_{XY}^{-1}[\ell_i, \ell_j] \nonumber \\
&&\times
C_{XY}(\ell_i)C_{XY}(\ell_j), \label{eq:SN_CXY}
\eeqa
where ${\rm Cov}_{XY}[\ell_i, \ell_i]$ represents the covariance matrix between two modes of $\ell_i$ and $\ell_j$.
We assume that all the observable fields follow Gaussian distribution.
This is reasonable since our primary focus is on large-scale modes ($\simgt1$ deg) for which the linear perturbation theory gives accurate results.
For Gaussian fields $X$ and $Y$, 
the covariant matrix is described as 
\beqa
{\rm Cov}_{XY}[\ell_i, \ell_j] 
&=& 
\frac{\delta_{ij}}{(2\ell_{i}+1)\Delta \ell f_{\rm sky}}
\Bigl[C_{{\rm obs}, XX}(\ell_{i})C_{{\rm obs},YY}(\ell_{i}) 
\nonumber \\
&&
\,\,\,\,\,\,\,\,\,\,
\,\,\,\,\,\,\,\,\,\,
\,\,\,\,\,\,\,\,\,\,
\,\,\,\,\,\,\,\,\,\,
+
C^2_{{\rm obs},XY}(\ell_{i})\Bigr], \label{eq:cov_CXY}
\eeqa
where 
$f_{\rm sky}$ is the observed sky fraction.  
We consider binned power spectra with a bin width of 
$\Delta\ell$.
In Eq.~(\ref{eq:cov_CXY}), the observed spectra of 
$C_{{\rm obs}, XX}$, $C_{{\rm obs}, YY}$ and $C_{{\rm obs}, XY}$
include both clustering signal and shot noise.
We consider three fields, $\delta_{s, {\rm 2D}}$, 
${\rm DM}_{\rm est}$, and $\delta^{i}_{g, \rm 2D}$,
and then calculated the observed spectra as follows:
\beqa
C_{{\rm obs},ss}(\ell) &=& C_{ss}(\ell)+\frac{1}{{\bar n}_{s, \rm 2D}} \\
C_{\rm obs, DM-DM}(\ell) &=& C_{\rm DM-DM}(\ell)+\frac{\sigma^{2}_{\rm DM, host}}{{\bar n}_{s, \rm 2D}}, \\
C_{{\rm obs}, gi-gj}(\ell) &=& \delta_{ij} \left[ C_{gg, i}(\ell) +
\frac{1}{{\bar n}_{g, \rm 2D}}\right], \\
C_{{\rm obs}, sg, i}(\ell) &=& C_{sg, i}(\ell), \\
C_{{\rm obs}, {\rm DM}-g, i}(\ell) &=& C_{{\rm DM}-g, i}(\ell).
\eeqa 
The definitions of $C_{ss}$, $C_{\rm DM-DM}$, $C_{sg, i}$, and $C_{{\rm DM}-g, i}$ are shown in Secs.~\ref{subsec:FRB-DM}
and \ref{subsec:FRB-galaxy}, and the galaxy spectrum is defined as
\beqa
C_{gg,i}(\ell)
&=& \int {\rm d}z\, W^2_{g,i}(z) \frac{H(z)}{\chi^2(z)} \nonumber \\
&&\, \times \,
b^2_{g, i} P_{m}\left(\frac{\ell+1/2}{\chi}, z\right).
\label{eq:C_gg}
\eeqa
Note that the shot noise is absent in the cross-correlation analysis.
In the following, we set the survey area to be 10,000 ${\rm deg}^2$, 
which roughly corresponds to the area of SDSS.
We set $\ell_{\rm max} = 200$, $\ell_{\rm min} = 10$ and $\Delta \ell = 50$ as constant 
and investigate the clustering signals of scales larger than $2\pi/\ell_{\rm max} = 1.8 \ \rm deg$.
Following results are not sensitive to the choice of $\Delta\ell$
since the power spectra have simple shapes as shown in Figures~\ref{fig:DM_llcl} and \ref{fig:cross_llcl}.

\subsubsection*{Autocorrelation of dispersion measures}

We first consider the autocorrelation of DMs. 
We here adopt our fiducial model of $C_{\rm DM-DM}(\ell)$ 
as shown in Fig.~\ref{fig:DM_llcl}.
We then examine the effect of shot noise 
on the detectability of the clustering signal.

The top panel in Fig.~\ref{fig:SN_DM_llcl} shows the S/N of $C_{\rm DM-DM}$ for various average source number densities.
For an intrinsic scatter of $\sigma_{\rm DM, host}=100\, {\rm pc} \ {\rm cm}^{-3}$,
the clustering signal can be identified with a 5$\sigma$ significance by detecting $\sim$ 1000 FRBs.  
Once 10,000 FRBs are observed, the S/N can be $\sim10$, 
corresponding to a measurement of $C_{\rm DM-DM}$ with a 
$\sim10\%$ accuracy.
The effect of $\sigma_{\rm DM, host}$ on the detectability of $C_{\rm DM-DM}$ is shown in the bottom panel 
where we assume 10,000 FRBs are observed in a 10,000 ${\rm deg}^2$ field. 
According to this figure, we can measure $C_{\rm DM-DM}$ with a $10-20\%$ accuracy
even if $\sigma_{\rm DM, host}$ is 
of an order of $100 \, \rm pc \ {\rm cm}^{-3}$.

\begin{figure*}
\begin{center}
       \includegraphics[clip, width=1.50\columnwidth, bb=0 0 564 329]
       {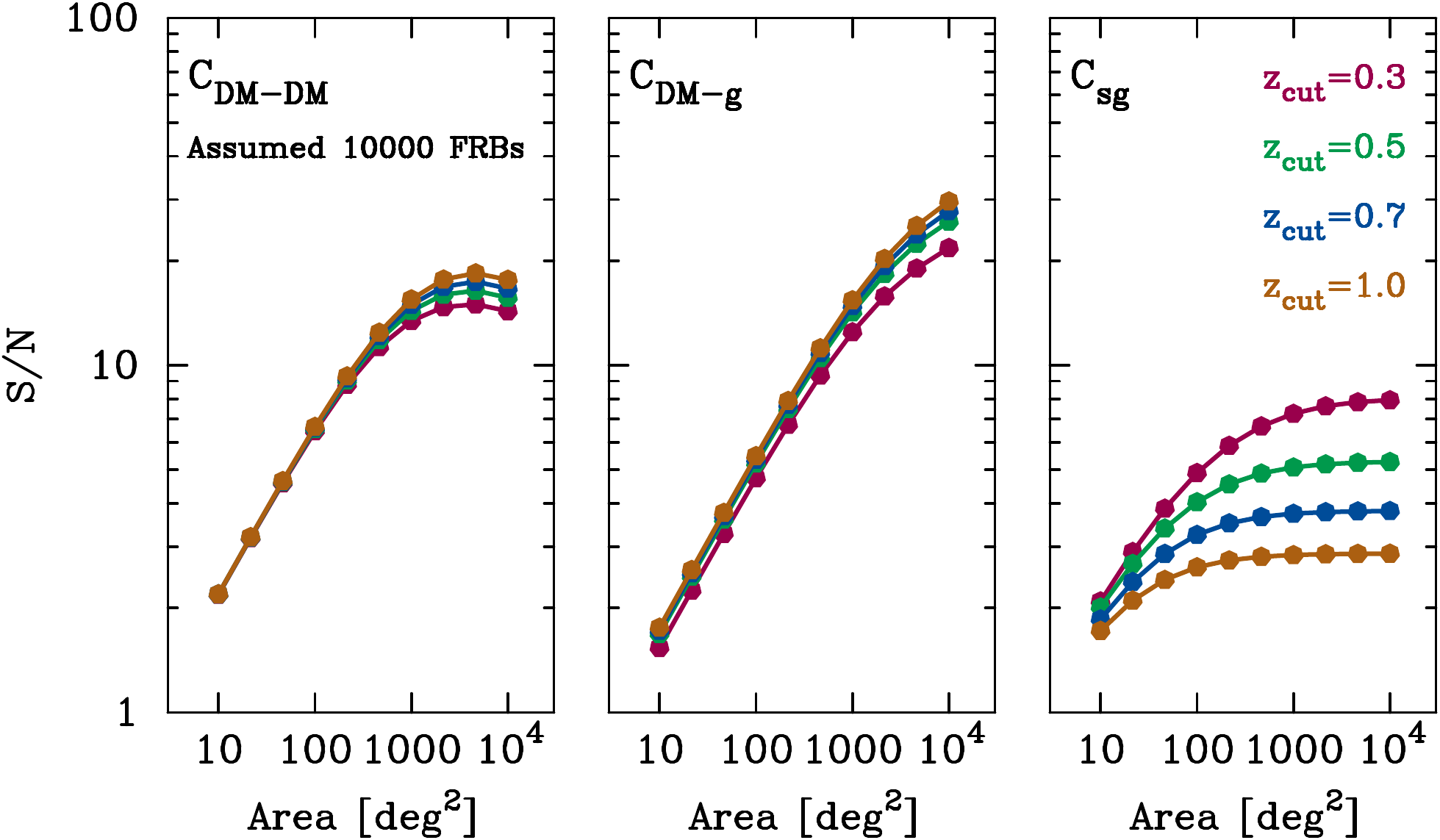}
   \caption{
     \label{fig:SN_fsky_zcut}
     Dependence of the ${\rm S/N}$ of FRB clustering signals on the survey parameters.
     In each panel, different color lines represent the different values 
     of redshift cutoff $z_{\rm cut}$  in Eq~(\ref{eq:ps_z}), 
     corresponding to different detection thresholds of radio survey. 
     We set the maximum multipole as $\ell_{\rm max} = 200$ 
     and the intrinsic scatter of DM as $\sigma_{\rm DM, host} = 100\, {\rm pc} \ {\rm cm}^{-3}$.
     The left, medium, and right panels show 
     the cases of autocorrelation of DMs ($C_{\rm DM-DM}$),
     cross-correlation of DMs and galaxies
     ($C_{\rm DM-g}$),
     and cross-correlation of positions between sources and galaxies
     ($C_{sg}$),
     respectively.
     }
    \end{center}
\end{figure*} 

\subsubsection*{Cross-correlation with galaxy distribution}

We next consider the cross-correlation with galaxy distribution.
When computing the S/N, we adopt the fiducial model as shown in Fig.~\ref{fig:cross_llcl}.
Fig.~\ref{fig:SN_cross_llcl} summarizes the S/N
of cross power spectra as a function of average source number density.

The upper panels show the cases of cross-correlation between 
FRB sources and galaxies, while the bottom panels
are for the correlation between DMs and galaxies.
Compared with the autocorrelation (Fig.~\ref{fig:SN_DM_llcl}), 
we require a larger number of events to detect the clustering signals;  
10,000 events are necessary to detect $C_{sg}$ for LOW-Z and CMASS samples with a $\sim3\sigma$ significance, 
whereas it is difficult to detect the signal for the highest redshift bin.
We also find that the shot noise will dominate for average source number density of $\bar n_{\rm s, 2D}<10 \, {\rm deg}^{-2}$.

The bottom panels show that the S/N of $C_{{\rm DM}-g}$ can be close to the cosmic-variance limit with 10,000 FRBs in the $10,000\, {\rm deg}^2$ field, 
resulting in the measurement with a $\sim5-10\%$ accuracy 
for LOW-Z and CMASS samples and $\sim50\%$ accuracy for eBOSS sample.
The effect of $\sigma_{\rm DM, host}$ on the detectability of $C_{\rm DM-host}$ is similar to the case of $C_{\rm DM-DM}$ (Fig.~\ref{fig:SN_DM_llcl} bottom).
The S/N will be degraded by a factor of $3-5$  when we increase $\sigma_{\rm DM, host}$ from $20 \, {\rm pc} \ {\rm cm}^{-3}$ to $100 \, {\rm pc} \ {\rm cm}^{-3}$.

We should note that a measurement of $C_{sg}$ with a 5\% accuracy roughly leads to constrain $b_{\rm FRB}W_s(z_{g})$ with a similar accuracy 
where $z_{g}$ represents the redshift of a given galaxy sample (see also Eq.~\ref{eq:C_sg}).
Likewise a measurement of $C_{{\rm DM}-g}$ with a 5\% accuracy constrains $b_{e}W_{\rm DM, IGM}(z_{g})$ with a $\sim$5\% accuracy.

\begin{table*}[!t]
\begin{center}
\begin{tabular}{|c|c|c|c|}
\tableline
Analysis& Parameters of interest & Physical meaning & Fiducial value \\ \tableline
$C_{\rm DM-DM}$, $C_{{\rm DM}-g}$ & $b_{e}$ & The fraction of free electrons in the unit of 0.88 & 1 \\ 
$C_{\rm DM-DM}$, $C_{{\rm DM}-g}$ & $b_{\rm FRB}{\bar \tau}_e$ & The bias of sources times mean DM around source population 
& $1.3\times100\, {\rm pc} \ {\rm cm}^{-3}$ \\
$C_{sg}$ & $b_{\rm FRB}$ & The bias of sources & 1.3 \\
$C_{\rm DM-DM}$, $C_{{\rm DM}-g}$, $C_{sg}$ & $W_s(z)$ & The redshift distribution of sources of FRBs & Eq~(\ref{eq:ps_z}) \\
\tableline
\end{tabular} 
\caption{
\label{tab:fisher}
Set of parameters that will be constrained 
by large-scale clustering of FRBs.
In principle, the autocorrelation of source number density field $C_{\rm ss}$
should also contain some information of $W_{s}(z)$ and $b_{\rm FRB}$. 
However, it is expected to be difficult to detect the signal from $C_{\rm ss}$ even with $\sim 10,000$ FRBs.
}
\end{center}
\end{table*}

\begin{figure*}
\begin{center}
       \includegraphics[clip, width=0.90\columnwidth, bb=0 0 470 480]
       {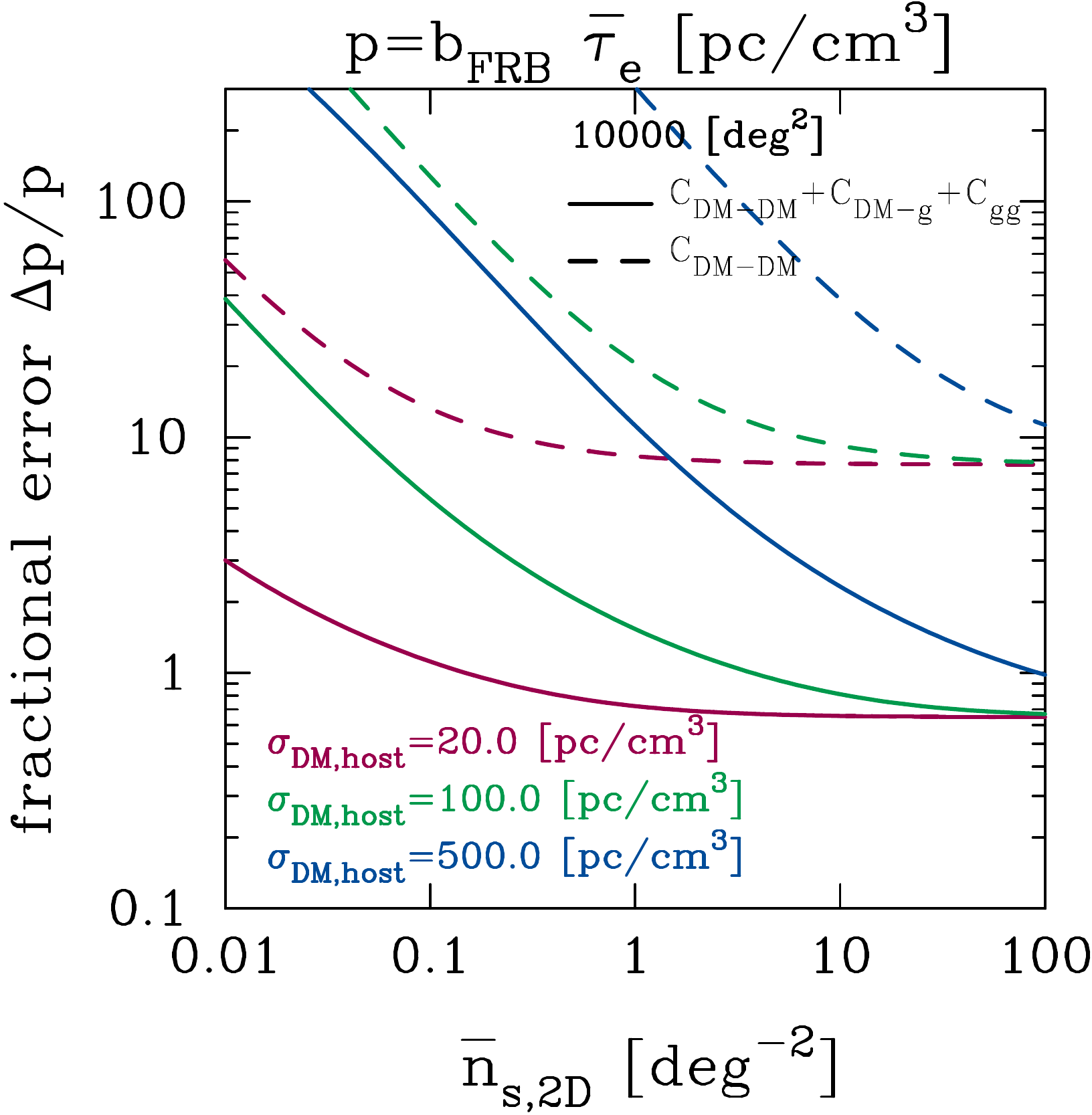}
       \includegraphics[clip, width=0.90\columnwidth, bb=0 0 470 480]
       {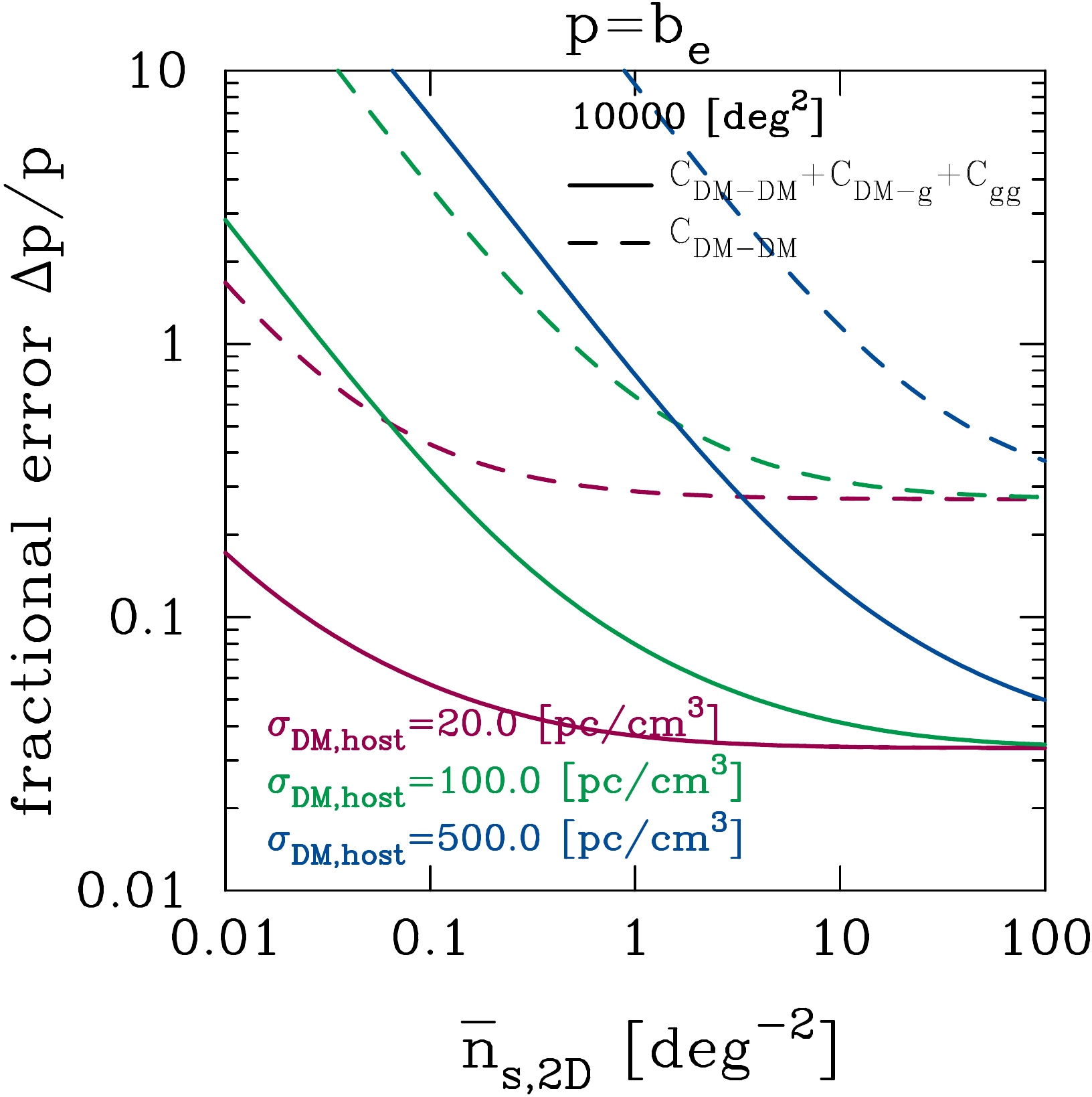}
     \caption{
     \label{fig:onesigma}
     Fractional errors of parameters
     as a function of average source number density and intrinsic
     scatter of DM around host galaxies $\sigma_{\rm DM, host}$.
     The left panel shows the case for 
     $b_{\rm FRB}{\bar \tau}_e$ and the right is for $b_e$.
     In both panels, the dashed lines represent the constraints
     only from the auto-correlation of DMs 
     and the solid lines are for 
     the combined analysis with the cross correlation 
     with galaxy distribution.
     The color difference corresponds to the difference of $\sigma_{\rm DM, host}$: $20\, {\rm pc} \ {\rm cm}^{-3}$ in red,
     $100\, {\rm pc} \ {\rm cm}^{-3}$ in green
     and 
     $500\, {\rm pc} \ {\rm cm}^{-3}$ in blue.
     }
    \end{center}
\end{figure*}

\subsubsection*{Dependence on the FRB survey configuration}

So far we consider a specific detection threshold 
in FRB clustering analyses.
In our theoretical framework shown in Sec.~\ref{sec:FRB}, 
the redshift cutoff $z_{\rm cut}$ in Eq~(\ref{eq:ps_z}) represents the detection threshold of the FRB survey; 
a smaller $z_{\rm cut}$ corresponds to a lower sensitivity. 
Radio surveys can be roughly categorized into two types:
(i) a low detection threshold with a large sky coverage
and
(ii) a high detection threshold with a small sky coverage.
The former corresponds to ``imaging" survey,
while the latter is ``beam-formed" survey.
Here we calculate the S/N of FRB clustering signals
as a function of the total survey area and $z_{\rm cut}$
in order to demonstrate which survey strategy 
(beam-formed or imaging) will be suitable.

Fig.~\ref{fig:SN_fsky_zcut} summarizes the S/N
of three clustering analyses $C_{\rm DM-DM}$, $C_{\rm DM-g}$,
and $C_{sg}$ as a function of the total area and $z_{\rm cut}$.
In this figure, we assume 10,000 detections 
and adopt the fiducial model of clustering signals. 
For cross-correlation analyses of $C_{\rm DM-g}$ and $C_{sg}$,
we properly combine three redshift bins of galaxies that are given by Table~\ref{tab:gal}.
In general, a larger sky coverage improves the S/N more efficiently 
and the effects of the cutoff $z_{\rm cut}$ are not significant. 
This is simply because the statistical uncertainty of the clustering analyses scales with inverse of survey area.
Although a hypothetical survey with a larger source number density in the same sky coverage can suppress the shot noise in clustering signals, 
the statistical uncertainty will always dominate.

The effects of $z_{\rm cut}$ will appear in the clustering signals of $C_{\rm DM-DM}$ and $C_{\rm DM-g}$
only when the sky coverage will be close to $\sim10,000$ squared degrees.
This is because that $C_{\rm DM-DM}$ and $C_{\rm DM-g}$ is mainly determined by low-redshift structures. 
According to the left and medium panels,
the S/N of $C_{\rm DM-DM}$ and $C_{\rm DM-g}$ will vary from $\sim 10$ to $\sim 20$ for $z_{\rm cut} = 0.3-1.0$. 
This suggests that the difference of $z_{\rm cut}$ will induce a $\sim1/20-1/10 \sim 5$\% effect on the clustering signals.
It should be noted that the S/Ns converge for $z_{\rm cut} \simgt0.5$.

\begin{figure}[!t]
\begin{center}
       \includegraphics[clip, width=0.9\columnwidth, bb=0 0 481 425]
       {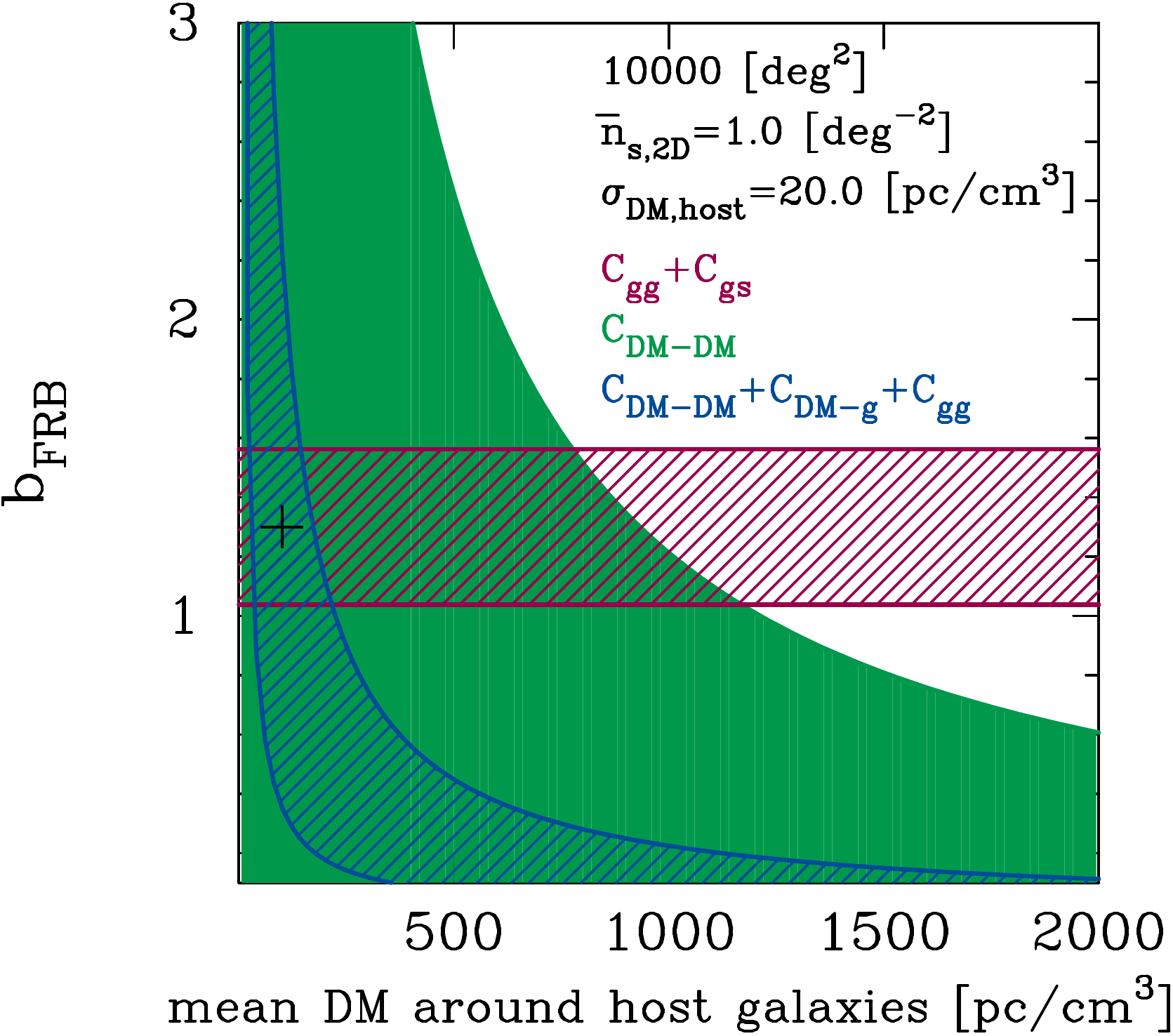}
     \caption{
     \label{fig:b_bmeanDM}
     The expected constraint in ${\bar \tau}_{e}-b_{\rm FRB}$ plane.
     The green filled region shows the 68\% confidence region from $C_{\rm DM-DM}$ alone,
     while the blue hatched region represents the 68\% confidence region 
     from the combined analysis with 
     $C_{gg}$, $C_{{\rm DM}-g}$ and $C_{\rm DM-DM}$.
     The red is for the combined analysis 
     with $C_{gg}$ and $C_{gs}$.
     In this figure, we assume 10,000 events with the sky coverage 
     of 10,000 ${\rm deg}^2$ and the intrinsic scatter of DM around host
     $\sigma_{\rm DM, host}=20\, {\rm pc} \ {\rm cm}^{-3}$.
     }
    \end{center}
\end{figure}

The S/N of $C_{sg}$ will be affected by $z_{\rm cut}$ when 
the sky coverage is larger than $\sim100$ squared degrees.
The statistical uncertainty of $C_{sg}$ is mainly determined
by the poisson term (source number density) 
when 10,000 detections are assumed.
Among the galaxy samples in Table~\ref{tab:gal},
LOW-Z and CMASS contribute the most 
to the signal of $C_{sg}$ since the structures at lower-redshift
have a larger clustering amplitude at degree scales.
We find that lowering $z_{\rm cut}$ can improve the S/N of $C_{sg}$ more efficiently.
For example, with a sky coverage of 10,000 square degrees,
the S/N of $C_{sg}$ with $z_{\rm cut}=0.3$ is larger than that with $z_{\rm cut}=1.0$ by a factor of $\sim3$.

From the above results, we conclude that 
future imaging surveys with $z_{\rm cut}\sim0.5$ 
and a larger sky coverage as possible are suitable 
for large-scale clustering analyses of FRBs.
   
\section{\label{sec:Fisher}Parameter constraints}

In Sec.~\ref{subsec:SNR}, we show that 
$\sim$ 10,000 FRBs over the sky coverage of $\sim$ 10,000 ${\rm deg}^2$
will enable to detect the large-scale clustering signals of FRBs with a high S/N.
Here we investigate what we can learn from such precise measurements.
We perform a Fisher analysis of the FRB autocorrelation 
and the cross-correlation with SDSS galaxies
in order to quantify the constraints on the model parameters. 
The details of our Fisher analysis are summarized in Appendix~\ref{app:fisher} and 
the parameters of interest are shown in Table~\ref{tab:fisher}\footnote{
It's worth mentioning that the forecast in the Fisher analysis depends on
the choice of fiducial value of the parameters in principle.
Nevertheless, the expected error in FRB parameters is found to be 
less affected by the choice of fiducial value 
if we set $\bar n_{\rm s, 2D}\simeq1\, {\rm deg}^{-1}$
and $\sigma_{\rm DM, host}\simeq100\, {\rm pc}\, {\rm cm}^{-3}$.
This is because the statistical uncertainty of clustering signals is mainly determined 
by the Poisson noise in our case
and the clustering signal is assumed be proportional to
the parameters of $b_{e}$, $b_{\rm FRB}$ and $b_{\rm FRB}\bar{\tau}_{e}$.
}.

Fig.~\ref{fig:onesigma} shows the dependence of the determination accuracy of $b_{\rm FRB}{\bar \tau}_e$ and $b_{e}$ 
with respect to $\bar n_{\rm s, 2D}$ and $\sigma_{\rm DM, host}$. 
We find that one can determine $b_{e}$ with an uncertainty of 30-1000\% by $C_{\rm DM-DM}$ alone with $\sim$10,000 FRBs (dashed lines in the right panel).
On the other hand, $C_{\rm DM-DM}$ will not put a meaningful constraint on $b_{\rm FRB}{\bar \tau}_{e}$ 
if the IGM contribution is dominant as in our fiducial model (dashed lines in the left panel).
Importantly, by combining the cross correlation $C_{{\rm DM}-g}$ and $C_{gg}$, 
the constraints on $b_{\rm FRB}{\bar \tau}_e$ and $b_{e}$ can be significantly improved by a factor of $\sim10$ for a wide range of $\bar n_{\rm s, 2D}$ and $\sigma_{\rm DM, host}$ (solid lines).

Only using $C_{\rm DM-DM}$, $C_{{\rm DM}-g}$ and $C_{gg}$, the constraints on $b_{\rm FRB}$ and ${\bar \tau}_{e}$ are strongly degenerate. 
This degeneracy is resolved by adding $C_{sg}$ 
as demonstrated in Fig.~\ref{fig:b_bmeanDM}. 
With 10,000 FRBs in the 10,000 $\rm deg^{2}$ field and $\sigma_{\rm DM, host} = 20\, {\rm pc} \ {\rm cm}^{-3}$,
the fractional error of $b_{\rm FRB}$ and ${\bar \tau}_e$ can become as small as $\sim 20\%$ and $\sim 70\%$, respectively. 
More generally, 
the fractional errors of $b_{\rm FRB}$ and 
$b_{\rm FRB} {\bar \tau}_e$
are approximately given as
\beqa
\frac{\Delta b_{\rm FRB}}{b_{\rm FRB}}
&\simeq& 
0.2 \left(\frac{{\bar n}_{s, \rm 2D}}{1\, {\rm deg}^{-2}}\right)^{-0.5}, \\ 
\label{eq:frac_err_b} 
\frac{\Delta (b_{\rm FRB} {\bar \tau}_e)}{b_{\rm FRB} {\bar \tau}_e}
&\simeq& 
0.7 x^2 \Big\{0.05 \left(\frac{\sigma_{\rm DM, host}}{20.0\, {\rm pc}\, {\rm cm}^{-3}}\right)^{-0.3} \nonumber \\
&& 
\,\,\,\,\,\,\,\,\,
\,\,\,\,\,\,\,\,\,
\,\,\,\,\,\,\,\,\,
\,\,\,\,\,\,\,\,\,
\,\,\,\,\,\,\,\,\,
\,\,\,\,\,\,\,\,\,
\,\,\,\,\,\,\,\,\,
+ 1/x^2 \Big\},
\label{eq:frac_err_btaue}
\eeqa
where $x$ in Eq.~(\ref{eq:frac_err_btaue}) is defined as
\beqa
x = \left(\frac{\sigma_{\rm DM, host}}{20.0\, {\rm pc}\, {\rm cm}^{-3}}\right)
\left(\frac{{\bar n}_{s, \rm 2D}}{1\, {\rm deg}^{-2}}\right)^{-1/2}.
\eeqa
Note that the bias factor of star-forming galaxies and passive galaxies at $z<1$ are $b_{\rm FRB}=1.3$ and $= 1.7-1.9$, respectively~\cite{Zehavi:2010bh}
and the difference is also $\sim 20 \ \%$. 
Thus, the host galaxy type of FRBs can be statistically inferred once $\sim$ 10,000 FRBs are detected in the 10,000 $\rm deg^{2}$ field. 

So far we have assumed that the FRB source distribution follows the star-formation history (Eqs.~\ref{eq:ps_z} and \ref{eq:SFH}). 
Since $C_{\rm DM-DM}$, $C_{{\rm DM}-g}$ and $C_{sg}$ contain information of $W_s$ [Eqs.~(\ref{eq:C_DMDM}), (\ref{eq:C_sg}), and (\ref{eq:C_DMg})], 
the redshift distribution of FRB sources can be also constrained from the clustering analysis, in principle. 
As a representative example, we here consider a time delay distribution
$f(\Delta t) \propto (\Delta t)^{-\alpha_t}$ with $\Delta t>20\, {\rm Myr}$.
For a given $\alpha_{t}$, we can compute the source distribution
by convoluting the delay time distribution
and the global star-formation history:
\beqa
{\bar n}_{s, \rm delay}(z; \alpha_{t})
&=& {\cal A} \exp\left[-\frac{d_{L}^{2}(z)}{2d_{L}^{2}(z_{\rm cut})}\right] \nonumber \\
&&\times
\int_{z}^{\infty}{\rm d}z^\prime\,
{\dot \rho}_{*}(z^{\prime}; \alpha_{1}=0.13) \nonumber \\
&&
\,\,\,\,\,\, 
\,\,\,\,\,\, 
\times \left(\frac{t(z)-t(z^\prime)}{\Delta t_{\rm norm}}\right)^{-\alpha_t} \frac{{\rm d}t}{{\rm d}z^{\prime}}
\label{eq:delay_model}
\eeqa
where $t(z)$ is the age of universe as a function of $z$,
$\Delta t_{\rm norm}$ is the normalization factor for the delay time distribution
and $\dot{\rho}_{*}(z)$ is given by Eq.~(\ref{eq:SFH}).
By comparing our fiducial model [Eq.~(\ref{eq:ps_z})] and ${\bar n}_{s, \rm delay}(z; \alpha_{t})$,
we find the approximated correspondence between two models:
$\alpha_{t}=0.5$ in Eq.~(\ref{eq:delay_model}) 
corresponds to $\alpha_{1}=0$ in Eq.~(\ref{eq:ps_z}),
while 
$\alpha_{t}=1.0$ in Eq.~(\ref{eq:delay_model}) 
is for $\alpha_{1}=0.2\times0.13$ in Eq.~(\ref{eq:ps_z}).

We find that the fractional error of $\alpha_{1}$ scales as
\beqa
\frac{\Delta {\alpha}_1}{{\alpha}_1}
\simeq 4.0 \left(\frac{{\bar n}_{s, \rm 2D}}{1\, {\rm deg}^{-2}}\right)^{-0.5}, \label{eq:frac_err_p1}
\eeqa
with $C_{sg} + C_{gg}$
for the sky coverage of 10,000 ${\rm deg}^{2}$.
Eq~(\ref{eq:frac_err_p1}) shows that we require $\simgt160,000$ events
on 10,000 ${\rm deg}^{2}$ to constrain on $\alpha_{t} < 0.5$, 
whereas $4\times10^{6}$ events enable us to constrain on $\alpha_{t}<1$.
It should be noted that the model of $\alpha_{t}=1$
roughly corresponds to the neutron-star merger scenario~\cite{1992ApJ...389L..45P}.

\begin{figure*}
\begin{center}
       \includegraphics[clip, width=0.91\columnwidth, bb=0 0 469 417]
       {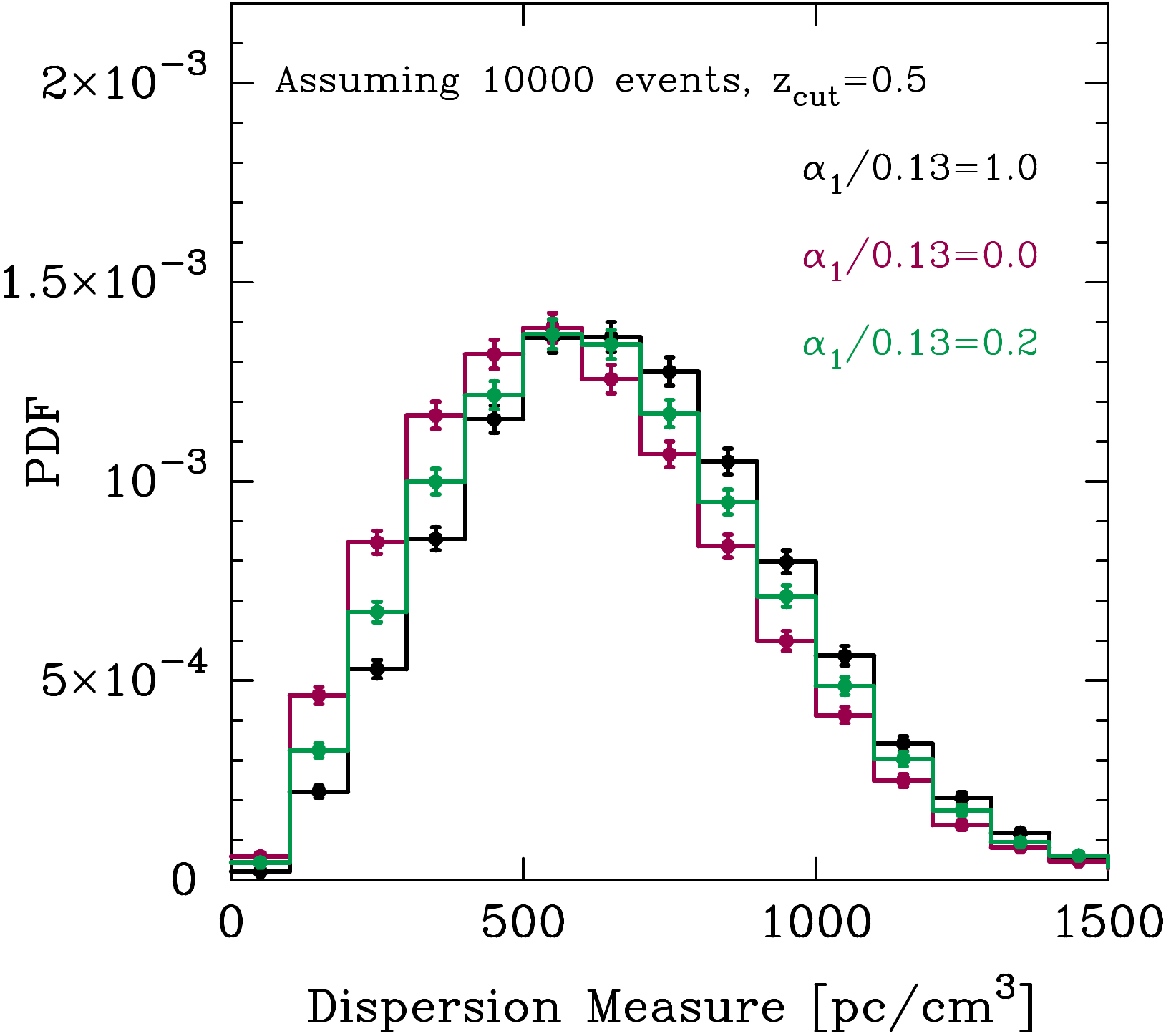}
       \includegraphics[clip, width=0.90\columnwidth, bb=0 0 460 417]
       {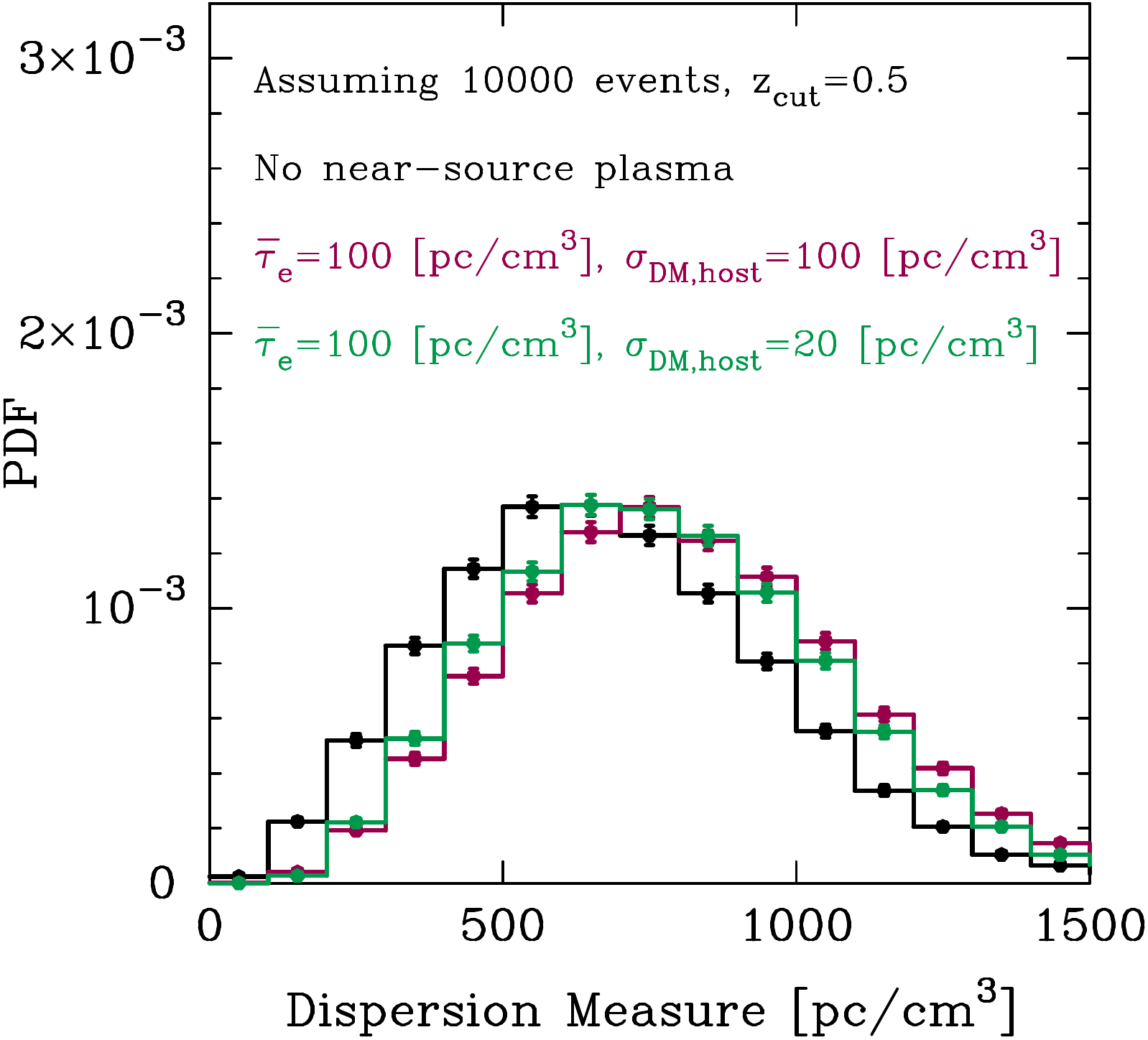}
     \caption{
     \label{fig:DM_dist}
     Dependence of the probability distribution function (PDF) of observed DM on FRB source distribution (left) and DM around host galaxies (right).
     In both panels, the black line shows our fiducial case with a source distribution of Eq.~(\ref{eq:ps_z}) and $\bar{\tau}_e = 0$.
     The colored lines in left panel represent the model PDF with different values of $\alpha_{1}$ [Eq.~(\ref{eq:ps_z})].
     The red and green lines in right panel show the cases with $\bar \tau_e = 100 \ \rm pc \ cm^{-3}$ and $\sigma_{\rm DM, host}=100\, {\rm pc}\, {\rm cm}^{-3}$ or $20\, {\rm pc}\, {\rm cm}^{-3}$, respectively. 
     The error bars in both panels show the poisson error for 10,000 events.
     }
    \end{center}
\end{figure*}

\subsection{\label{subsec:DM_dist}Combining DM distribution function}

As shown in the previous section, even 10,000 FRBs are not enough 
to obtain meaningful constraints on the redshift distribution of FRB sources 
if we only uses the large-scale clustering. 
In order to improve the constraint, we need additional information
other than two-point correlation functions.

One of the simplest FRB statistics is one-point distribution
function or probability distribution function (PDF) of DM. 
It is suggested that the DM PDF contains rich cosmological information \cite{McQuinn:2013tmc}. 
For example, Ref. \cite{Dolag:2014bca} proposes to use the DM PDF
to determine the formation mechanism of FRBs.
Here we explore how the constraint on the redshift distribution
of FRB sources can be improved by combining the DM PDF.

The left panel of Fig.~\ref{fig:DM_dist} shows the DM PDFs with
different source redshift distributions\footnote{
The details of our modeling of cosmological DM are summarized in Appendix~\ref{app:N-body}.}.
The black line shows our fiducial model 
with $\alpha_{1}=0.13$ in Eq.~(\ref{eq:ps_z}),
while the red and green lines are for $\alpha_{1}=0$ 
and $0.13\times0.2$, respectively.
As expected, the mean value of DM becomes smaller for a larger delay time. 
The error bar in Fig.~\ref{fig:DM_dist} represents the poisson error for 10,000 FRBs.
The left panel clearly shows the statistical power of DM PDF to constrain
the source redshift distribution.
As in Sec.~\ref{sec:Fisher}, the model of $\alpha_{t}=1$
roughly corresponds to the neutron-star merger scenario;
the DM PDF of 10000 events is sufficient
to constrain the various time-delayed models in Eq~(\ref{eq:delay_model}).
However, we should stress that in the left panel we neglect the contribution from ${\rm DM}_{\rm host}$, 
which in general affect the observed DM PDF. 

The right panel in Fig.~\ref{fig:DM_dist} shows the impact of ${\rm DM}_{\rm host}$ on DM PDF.
As for the PDF of $\rm DM_{\rm host}$,
we assume that ${\rm DM}_{\rm host}$ follows Gaussian distribution with 
mean of $\bar{\tau}_{e}$ and scatter of $\sigma_{\rm DM, host}$.
The black line in the right panel is the same as that in the left panel 
while the red and green lines correspond to the cases
with $\bar{\tau}_e = 100\, {\rm pc}\, {\rm cm}^{-3}$
and $\sigma_{\rm DM, host}=100\, {\rm pc}\, {\rm cm}^{-3}$ or $20\, {\rm pc}\, {\rm cm}^{-3}$, respectively.
In the right panel, we set the source distribution as our fiducial model [Eq.~(\ref{eq:ps_z})].
Of course, the mean value of $\rm DM$ becomes larger when including ${\rm DM}_{\rm host}$. 
One can see that the effects of $\sigma_{\rm DM, host}$ will be minor as far as $\sigma_{\rm DM, host} \lesssim \bar{\tau}_e$. 

Fig.~\ref{fig:DM_dist} shows that
DM PDF is a powerful probe of the redshift distribution of FRB sources.
Note, however, that expected constraints from DM PDF are dependent on
the intrinsic properties of ${\rm DM}_{\rm host}$.
Unfortunately, our result indicates that it is difficult 
to determine the source distribution
and the mean host DM with PDF alone.
In contrast, the large-scale clustering of FRBs has a good sensitivity
for $\bar{\tau}_{e}$ and $\sigma_{\rm DM, host}$.
Therefore, by combining the large-scale clustering and PDF of FRBs,
the redshift distribution of FRB sources can be also constrained with a similar accuracy to $\bar{\tau}_{e}$ in Figure \ref{fig:b_bmeanDM}. 
In order to study more detailed information content in the combined
analysis, we require more accurate modeling of FRBs
and leave it for our future work.
     
\section{\label{sec:con}CONCLUSION AND DISCUSSION}

In this paper, we have studied the information content in 
large-scale clustering of FRBs at degree scales.
We have developed a theoretical framework for the clustering analyses
based on the standard theory of structure formation.
In addition to the two-point clustering of FRB source number density
and extragalactic DMs, we have considered 
the cross-correlation with
galaxy distributions to identify the origin of FRBs.
Assuming a reasonable parameter set, 
we have investigated the S/N of clustering signals and made a forecast for 
expected constraints on the model parameters obtained by future radio transient surveys.
Our main findings are summarized as follows:

\begin{enumerate}

\vspace{2mm}
\item 
The autocorrelation of DMs consists of contributions
from the clustering of IGM, the clustering of host galaxies, 
the clustering due to overlapped redshift distribution between IGM
and host galaxies, and the shot noise originating from the intrinsic scatter
of DM around host galaxies.
Among these, the IGM clustering is likely to be dominant in 
the autocorrelation of ${\rm DM}_{\rm ext}$.
The typical amplitude is expected to be 
$\sim$0.1-10 $\, [{\rm pc} \ {\rm cm}^{-3}]^2$
in the range of $\ell\sim$10-200.
The clustering between IGM and host galaxies can be significant
if the mean DM around host galaxies ${\bar \tau}_e$ is
$\sim$ 600-700 $\, {\rm pc} \ {\rm cm}^{-3}$.

\vspace{2mm}
\item 
The S/N of autocorrelation of DMs depends
on the average source number density ${\bar n}_{s, \rm 2D}$
and the intrinsic scatter of DM
around host galaxies $\sigma_{\rm DM, host}$
for a fixed survey area.
Assuming a hypothetical survey with the sky coverage of 
$10,000\, {\rm deg}^{2}$ and 
$\sigma_{\rm DM, host}=100\, {\rm pc} \ {\rm cm}^{-3}$, 
we estimate that 1,000 events are sufficient 
to detect the clustering signal of IGM with a $3\sigma$ significance.
A sample of 10,000 FRBs enable us to measure the signal 
with a $\sim10\%$ accuracy at degree scales.
A similar S/N can be obtained
in the cross-correlation of DMs and the galaxy distribution 
from existing spectroscopic galaxy samples.
The cross-correlation of FRBs
with galaxy distributions in the redshift range of $0.15<z<1.6$
can be detected with a $\simgt3 \sigma$ confidence level 
if $\sim$ 10,000 FRBs are observed.

\vspace{2mm}
\item 
Measurement of large-scale clustering of FRBs can place
constraints on the fraction of free elections,
the environment of the source population(s),
and the mean DM around host galaxies.
The DM autocorrelation can be used to constrain 
the global abundance of free electrons at $z<1$
with a level of $\sim70\%$,
if 10,000 FRBs are observed over $10,000 \,{\rm deg}^{2}$
and the intrinsic scatter of DM is assumed to be 
$\sigma_{\rm DM, host}=100\, {\rm pc} \ {\rm cm}^{-3}$.
The cross-correlation with galaxy distributions 
will improve the constraint by a factor of $\sim10$.
The cross-correlation of FRBs and galaxy distributions
will help determining 
the linear bias of the source population $b_{\rm FRB}$
with a level of $\sim20\%$.
If we add the information 
from the DM-galaxy cross-correlation,
it is possible to put a tight constraint on 
the mean DM around host galaxies 
by statistical analysis in future transient surveys
(see Fig.~\ref{fig:b_bmeanDM}).



\end{enumerate}

Our clustering analysis can be useful to identify the origin of FRB. 
In some models, 
FRBs are associated with newborn or young compact stellar objects, 
e.g, fast-spinning pulsars or magnetars~\citep{Popov&Postnov10,Connor+16,Lyuvarsky14}. 
In this case, FRBs typically occur in star-forming galaxies and the bias factor 
will be $b_{\rm FRB} \sim 1.3$. 
The first identified FRB host of FRB 121102 may belong to this group~\citep{Tendulkar_et_al_17}. 
On the other hand, e.g., in the compact binary merger scenarios~\citep{Kashiyama+13,Totani13}, 
FRBs will preferentially occur in more evolved galaxies and 
the bias factor can range from $b_{\rm FRB} = 1.7-1.9$. 
In a more exotic scenario, e.g., evaporation of primordial black holes~\citep{Keane+12}, 
the bias factor could be $b_{\rm FRB} \sim 1$. 
Such a difference of the bias factor can be distinguished by the clustering analysis 
once $\sim$10,000 of FRBs are detected in a sky area of $\sim10,000\, {\rm deg}^2$. 
Another key to distinguish the FRB source candidates is the delay time distribution, 
which can be also constrained by the combined analysis of clustering and DM distribution function (see Fig.~\ref{fig:DM_dist}).
Although high-precision localization of FRBs with 
long-baseline observatories is still the most robust way 
to probe physical properties of FRB host galaxies and near source regions, 
a drawback is the small detection efficiency due to the limited field-of-view.  
Our statistical approach only requires an angular resolution of $\sim$ deg and will be complementary and powerful once $\sim 100-1000$ of FRBs are detected annually.

\begin{acknowledgements}
M.~S. is supported by Research Fellowships of the Japan Society for 
the Promotion of Science (JSPS) for Young Scientists.
N.~Y. and K.~K. acknowledge financial support from JST CREST.
Numerical computations presented in this paper were in part carried out
on the general-purpose PC farm at the Center for Computational Astrophysics,
CfCA, of the National Astronomical Observatory of Japan.
\end{acknowledgements}

\appendix
\section{Fisher Analysis}
\label{app:fisher}
Let us briefly summarize the Fisher analysis. 
For a multivariate Gaussian likelihood, the Fisher matrix
$F_{ij}$ can be written as
\beqa
F_{ij} = \frac{1}{2}{\rm Tr}\left[A_{i}A_{j}+C^{-1}H_{ij}\right],
\label{eq:fisher_matrix}
\eeqa
where
$A_{i}=C^{-1}\partial C/\partial p_i$,
$H_{ij} = 2(\partial \mu/\partial p_i)(\partial \mu/\partial p_j)$,
$C$ is the data covariance matrix,
$\mu$ represents the assumed model, and
$p_i$ describes parameters of interest. 
The Fisher matrix provides an estimate of the error covariance
for two parameters as
\beqa
\langle \Delta p_{\alpha} \Delta p_{\beta} \rangle = (F^{-1})_{\alpha \beta},
\eeqa
where $\Delta p_{\alpha}$ represents the statistical uncertainty
of parameter $p_{\alpha}$.

In the present study, we consider only the second term in 
Eq.~(\ref{eq:fisher_matrix}).  
Because $C$ is expected
to scale inversely to the survey area,  
the second term will be dominant for a large area survey.
We consider the following parameters to vary:
${\bd p}=\{b_{g, 1}, b_{g, 2}, b_{g, 3}, b_{\rm FRB}, {\alpha}_{1}, b_{\rm FRB}{\bar \tau}_{e}, b_{e}\}$
where $b_{g, i}$ is the galaxy bias for $i$ th spectroscopic sample
given by Table~\ref{tab:gal}
and ${\alpha}_{1}$ controls the redshift dependence of $W_{s}(z)$
[also see Eq~(\ref{eq:SFH})].
The fiducial values of ${\bd p}$ are set to be
${\bd p}_{\rm fid}=\{1.7, 1.9, 1.3, 1.3, 0.13, 
130\, {\rm pc} \ {\rm cm}^{-3}, 1\}$.

We construct the data vector $\bd D$ from a set of 
binned spectra $C_{\rm DM-DM}, C_{{\rm DM}-g}, C_{gg}$
and $C_{gs}$ as
\beqa
D_{i} &=& \{
C_{\rm DM-DM}(\ell_{1}),...,C_{\rm DM-DM}(\ell_{10}), 
\nonumber \\
&&
C_{{\rm DM}-g, 1}(\ell_{1}),...,C_{{\rm DM}-g, 1}(\ell_{10}),
\nonumber \\ 
&&
C_{{\rm DM}-g, 2}(\ell_{1}),...,C_{{\rm DM}-g, 2}(\ell_{10}),
\nonumber \\
&&
C_{{\rm DM}-g, 3}(\ell_{1}),...,C_{{\rm DM}-g, 3}(\ell_{10}),
\nonumber \\
&&
C_{gg, 1}(\ell_{1}),...,C_{gg, 1}(\ell_{10}),
\nonumber \\
&&
C_{gg, 2}(\ell_{1}),...,C_{gg, 2}(\ell_{10}),
\nonumber \\
&&
C_{gg, 3}(\ell_{1}),...,C_{gg, 3}(\ell_{10}),
\nonumber \\
&&
C_{gs, 1}(\ell_{1}),...,C_{gs, 1}(\ell_{10}),
\nonumber \\
&&
C_{gs, 2}(\ell_{1}),...,C_{gs, 2}(\ell_{10}),
\nonumber \\
&&
C_{gs, 3}(\ell_{1}),...,C_{gs, 3}(\ell_{10})
\},
\eeqa
where $\ell_{i} = \ell_{\rm min} + (i+0.5)(\ell_{\rm max}-\ell_{\rm min})/10$ with $\ell_{\rm min}=10$ and $\ell_{\rm min}=200$.
The cross covariance between two spectra of 
$C_{XY}$ and $C_{AB}$ is then computed as
\beqa
{\rm Cov}_{XY, AB}[\ell_{i}, \ell_{j}]
&=& \frac{\delta_{ij}}{(2\ell_{i}+1)\Delta\ell f_{\rm sky}}
\Bigl[C_{{\rm obs},XY}(\ell_{i})C_{{\rm obs},AB}(\ell_{i}) \nonumber \\
&&+C_{{\rm obs},XA}(\ell_{i})C_{{\rm obs},YB}(\ell_{i}) 
\Bigr],
\eeqa
where the width is set to be $\Delta \ell = (\ell_{\rm max}-\ell_{\rm min})/10$ and we assume the sky fraction of $f_{\rm sky}\simeq10000/41252.96=0.242$.

\section{Construction of mock FRB catalogs with cosmological $N$-body simulation}
\label{app:N-body}

Here we summarize our modeling of ${\rm DM}_{\rm IGM}$
based on cosmological $N$-body simulations.
For ${\rm DM}_{\rm IGM}$, we assume that 
the free-electron number density is 
an unbiased tracer of underlying matter density.
In order to simulate the three-dimensional matter density distribution, 
we utilize a set of $N$-body simulations used in 
our previous work of Ref~\cite{2012ApJ...760...45S}.
We employ $256^3$ particles in 
a comoving volume of $240^3\, [h^{-1}\, {\rm Mpc}]^3$
and damp ten snapshots in the redshift range of $z=0-1$.
We determine the output redshifts of simulation 
so that the simulation boxes are placed to 
cover a past light cone of a hypothetical observer with 
angular extent $5\times5\, {\rm deg}^2$
from redshift $z=0$ to $z\sim1$.
The details of our simulation are found in 
Ref~\cite{2012ApJ...760...45S}.

 From the distribution of $N$-body particles in each snapshot,
we first generate three-dimensional matter density field
on $256^3$ grids by using the nearest-grid-point method.
We then combine 10 grid-based density maps to generate
a light cone outout with a line-of-sight depth of $\sim2\, {\rm Gpc}$.
To avoid the same structure appearing multiple times
along the line of sight, 
we randomly shift the simulation boxes. 
In total, we generate 20 quasi-independent realizations 
of matter density distribution in a comoving 
volume of $240\times240\times2400 \, [h^{-1}\, {\rm Mpc}]^3$.
Note that the transverse grid size in the density maps
corresponds to a few arcmin 
at $z\sim0.5$. This is sufficient for 
upcoming wide-area FRB surveys such as CHIME.

We also locate dark matter halos using the standard
friend-of-friend (FOF) algorithm with the linking parameter of $b=0.2$.
We then assume that FRBs occur in 
dark matter halos with the FOF mass greater
than $10^{13}\, h^{-1}M_{\odot}$.
Note that the mass selection of $>10^{13}\, h^{-1}M_{\odot}$
roughly correspond to a sample with the halo bias of 1-1.5 \cite{2010ApJ...724..878T}.
Finally, we make a random downsampling of halos
so that the redshift distribution of FRB hosts can be approximated 
as assumed in our model.
For the input redshift distribution of FRBs, 
we consider the functional form of Eq.~(\ref{eq:ps_z})
with $z_{\rm cut}=0.5$.
As our fiducial model, we set $\alpha_{0}=0.0170$,
$\alpha_{1}=0.13$, $\alpha_{2}=3.3$, and $\alpha_{3}=5.3$,
while we examine a sensitivity of $\alpha_{1}$ on DM PDF.
After the random sampling,
we find $\sim2,000$ halos in each realization.
For selected halos, we compute ${\rm DM}_{\rm IGM}$
by summing the pixel value of grid-based matter density maps
along the line of sight as in Eq.~(\ref{eq:def_DM_IGM}),
assuming $\Omega_{b}h^2 = 0.022$ and $f_e=0.88$.

\bibliography{ref_prd}
\end{document}